\documentclass[11pt]{article}
\usepackage{draft}
\usepackage{xcolor}
\usepackage{cite}
\usepackage{cancel}
\usepackage{caption}  
\usepackage{graphicx} 
\usepackage{float} 
\usepackage{cite}
\usepackage{relsize}
\usepackage{physics}
\usepackage{psfrag}
\usepackage{cancel}
\usepackage{array}
\usepackage{amssymb}
\usepackage{amsmath, mathrsfs}
\usepackage{cite}
\usepackage{amsthm}
\usepackage{float}
\usepackage{tikz}
\usetikzlibrary{decorations.markings}
\usepackage{tikz,lipsum,lmodern}
\usepackage[most]{tcolorbox}
\usepackage{hyperref}
\usepackage{xcolor}
\usepackage{comment}
\usepackage{mathtools}
\usepackage{amsmath,amssymb,mathrsfs}
\usepackage{soul}
\usepackage{float}
\usepackage{enumerate}
\usepackage{cleveref}
\usepackage{booktabs}
\usepackage{array}

\usepackage{hyperref}

\usepackage{cleveref}

\crefformat{section}{\S#2#1#3} 
\crefformat{subsection}{\S#2#1#3}
\crefformat{subsubsection}{\S#2#1#3}

\usepackage{verbatim}

\usepackage[mathscr]{euscript}

\definecolor{twilightlavender}{rgb}{0.54, 0.29, 0.42}
\definecolor{richmaroon}{rgb}{0.69, 0.19, 0.38}
\definecolor{forestgreen(web)}{rgb}{0.13, 0.55, 0.13}
\definecolor{lava}{rgb}{0.81, 0.06, 0.13}
\hypersetup{
	breaklinks,
	colorlinks,
	citecolor=forestgreen(web),
	filecolor=richmaroon,
	linkcolor=lava,
	urlcolor=violet
}

\def\ee{\end{equation}}
\def\be{\begin{equation}}
\def\bea{\begin{eqnarray}}
\def\eea{\end{eqnarray}}
\newcommand{\beq}{\begin{eqnarray}}
\newcommand{\eqq}{\end{eqnarray}}
 \newcommand{\badat}{\begin{alignedat}}
 \newcommand{\eadat}{\end{alignedat}}

\newcommand{\eal}[1]{\be \begin{aligned} #1 \end{aligned}\end{equation}} 
\newcommand{\eqn}[1]{\be #1 \end{equation}} 
\newcommand{\eqa}[1]{\bea  #1\end{eqnarray}}

\renewcommand{\d}{\delta}
\newcommand{\til}[1]{\widetilde{#1}}

\newcommand{\bos}{\boldsymbol}
\newcommand{\bom}{\boldsymbol{\omega}}

\newcommand{\bost}{\bos{\theta}}
\newcommand{\ed}{\mathrm{d}}

\long\def\new#1\endnew{{\bf #1}}		
\long\def\del#1\enddel{}
\def\nn{\nonumber}

\usepackage{color}

\newcommand{\pink}[1]{\textcolor{\pink}{#1}}
\definecolor{dblue}{rgb}{0.2,0.50,0.80}

\def\S{\mathcal{S}}
\def\P{\mathcal{P}}

\def\A{\mathcal{A}}
\def\B{\mathscr{B}}
\def\C{\mathcal{C}}

\def\H{\mathcal{H}}
\def\cg{\mathcal{G}}
\def\M{\mathcal{M}}

\def\L{\mathcal{L}}

\def\F{\mathcal{F}}
\def\E{\mathcal{E}}
\def\Q{\mathcal{Q}}
\def\T{\mathcal{T}}

\def\ff{\mathfrak{F}}

\def\sf{\mathscr{F}}

\def\sc{\mathsf{C}}
\def\tf{\mathtt{F}}

\def\g{{\gamma}}

\def\a{\alpha}
\def\b{\beta}
\def\m{\mu}
\def\n{\nu}
\def\t{\tau}
\def\l{\lambda}

\def\veps{\varepsilon}
\def\pa{\partial}

\def\w{\wedge}
\def\cw{\curlywedge}

\def\hd{\hat{\ed}}
\def\hdel{\hat{\partial}}

\def\nn{\nonumber}

\def\sp{ \sigma}

\definecolor{richmaroon}{rgb}{0.69, 0.19, 0.38}

\newcommand{\pkd}[1]{\textcolor{forestgreen(web)}{\textsf{[Pratik: #1]}}}

\newcommand{\sd}[1]{\textcolor{richmaroon}{\textsf{[Sarthak: #1]}}}

\title{Covariant phase space approach to noncommutativity in tensile and tensionless open strings}

\affiliation[a]{Department of Physics, Indian Institute of Technology Indore,
Simrol, Khandwa Road, 453552, Indore, India}
\affiliation[b]{Yau Mathematical Sciences Center (YMSC), Tsinghua University, Beijing 100084, China}
\affiliation[c]{Harish-Chandra Research Institute, A CI of Homi Bhabha National Institute,
Chhatnag Road, Jhunsi, Prayagraj (Allahabad), Uttar Pradesh 211019, India}
\usepackage{orcidlink}
\author[a,\orcidlink{0009-0004-6269-3292}]{Pratik K. Das,}\emailAdd{ dask.pratik@gmail.com}
\author[b,\orcidlink{0000-0002-4535-3198}]{Sarthak Duary}\emailAdd{sarthakduary@tsinghua.edu.cn}
\author[c,\orcidlink{0009-0008-2963-2497}]{and Sourav Maji}\emailAdd{souravmaji@hri.res.in}

 \abstract{We study noncommutativity in open strings using the covariant phase space formalism. For tensile open strings in a constant Kalb-Ramond background, we show that the (pre)-symplectic current splits into a bulk kinetic term plus an exact boundary term, recovering the Seiberg-Witten noncommutativity parameter. We then extend the analysis to intrinsically tensionless strings. In the absence of background fields, the reduced phase space is degenerate and carries no intrinsic Poisson structure. In the presence of a constant Kalb-Ramond field, the symplectic current localises entirely on the boundary, so that the physical phase space becomes purely boundary-supported and the endpoint coordinates acquire a noncommutative Poisson algebra. Including a boundary gauge-field coupling similarly leads to a boundary symplectic form governed by the effective Born-Infeld combination on the D-brane. Our results provide a unified description of noncommutativity in both tensile and tensionless open strings.}

\begin{document}
\maketitle
\flushbottom

\section{Introduction}

The study of noncommutative geometry in string theory originated in the seminal works of Witten~\cite{Witten:1985cc} and, more directly, Seiberg and Witten~\cite{Seiberg:1999vs}, who showed that open strings propagating in a constant Kalb--Ramond background develop a nontrivial noncommutative structure at their endpoints. In the standard tensile theory, the background antisymmetric two-form modifies the boundary conditions obeyed by the open-string embedding coordinates on the D-brane. This deformation of the boundary dynamics is reflected in the worldsheet operator algebra: by analysing the boundary limit of the open-string two-point function, one finds that the endpoint coordinates fail to commute, and the resulting noncommutativity is encoded in the antisymmetric tensor \(\Theta^{\mu\nu}\), now known as the Seiberg--Witten parameter. This observation provided one of the clearest links between string dynamics in background fields and noncommutative geometry. It has since been developed and generalised in many directions, including mixed boundary conditions, D-brane bound states, noncommutative Yang--Mills theory, and the interpretation of boundary conditions as constraints~\cite{Sheikh-Jabbari:1997qke, Ardalan:1998ks, Sheikh-Jabbari:1998aur, Ardalan:1998ce, Sheikh-Jabbari:1999krr}.

Early work on Yang--Mills theory on noncommutative tori and its appearance in matrix-theory compactifications already suggested a deep relation between noncommuting coordinates, D-branes, and string dualities \cite{Connes:1997cr, Douglas:1997fm}. The perspective was further sharpened by the observation that T-duality acts naturally within the noncommutative Yang--Mills framework and is closely tied to Morita equivalence of noncommutative tori \cite{Schwarz:1998qj, Rieffel:1998vs, Astashkevich:1998uc, Morariu:1998qm, Brace:1998ai, Brace:1998ku, Brace:1998xz, Hofman:1998iy}. In addition, the study of instantons on noncommutative \(\mathbb{R}^4\) provided strong evidence that noncommutative gauge theory captures genuine D-brane physics in background \(B\)-fields, in particular through the resolution of small-instanton singularities and their interpretation in terms of modified ADHM data \cite{Nekrasov:1998ss, Berkooz:1998st}. These developments motivate treating noncommutative gauge theory not merely as a formal deformation, but as an intrinsic and efficient language for describing the low-energy open-string dynamics in suitable \(B\)-field backgrounds.

At the technical level, the standard derivation of this result is crucially dependent on structures that are special to the tensile string. The operator approach relies on the existence of a non-degenerate worldsheet metric, a well-defined two-dimensional conformal field theory (CFT), and the corresponding logarithmic propagator for the embedding fields. These ingredients enable separating the holomorphic and antiholomorphic sectors, explicitly solving the mixed boundary conditions, and extracting the endpoint commutator from the antisymmetric part of the boundary correlator. The tensionless limit \(T\to 0\), on the contrary, presents a qualitatively different regime. In this limit, the worldsheet geometry degenerates into Carrollian geometry, the conventional left- and right-moving decompositions cease to exist in their usual forms, and the theory becomes ultralocal in the spatial direction. Consequently, the standard oscillator expansion and the familiar logarithmic propagator are no longer available, so the usual route to derive endpoint noncommutativity breaks down. For this reason, tensionless theory calls for a framework that does not presuppose an ordinary conformal worldsheet description; see Section \ref{motdetails} for a more detailed discussion.

The covariant phase space (CPS) formalism provides precisely such a framework. Rather than deriving the phase space structure indirectly from operator algebras or propagators, the CPS method constructs it directly from the variational principle. Starting from the classical action, one identifies the symplectic potential from its first variation and the associated pre-symplectic current from a second antisymmetrised variation, thereby obtaining the symplectic structure on the space of classical solutions in a manifestly covariant manner. Developed through the foundational work of Zuckerman~\cite{Zuckerman:1986vzu}, Crnković and Witten~\cite{Crnkovic:1986ex, Crnkovic:1987tz}, and later refined by Lee, Iyer, and Wald~\cite{Lee:1990nz, Wald:1993nt, Iyer:1994ys}, this formalism offers a geometric description of phase space that remains meaningful even when the conventional canonical or operator-based approach becomes difficult to implement. Its more recent extension to systems with boundaries by Harlow and Wu~\cite{Harlow:2019yfa} is especially important for the present problem. In particular, the boundary-sensitive version of CPS is well suited to analyse how the antisymmetric \(B\)-field and additional gauge-field couplings contribute to the symplectic structure and therefore to derive the corresponding noncommutativity parameter in both tensile and tensionless regimes within a single covariant framework.

The CPS method has been widely used to understand gravitational charges and symmetries \cite{ Ashtekar:1990gc, Wald:1993nt, Iyer:1994ys, Wald:1999wa, Barnich:2007bf, Chandrasekaran:2018aop, Chandrasekaran:2021vyu, Adami:2021nnf, Ciambelli:2021nmv, Ciambelli:2022cfr, Ciambelli:2022vot}, the stability of black holes and black branes \cite{Hollands:2012sf}, in defining the entropy for dynamical black holes \cite{Wall:2015raa, Hollands:2024vbe, Visser:2024pwz} among the many other works in gravity. In \cite{Chandrasekaran:2026pnc}, it was shown that for null gravitational subregions, the standard CPS prescription is incomplete unless one includes corner edge modes.\footnote{The role of edge modes in subregion holography has also been explored via the CPS formalism \cite{Donnelly:2016auv, Speranza:2017gxd}.} These extra corner degrees of freedom supply the missing contribution to the subregion symplectic form, make half-sided symmetry generators integrable, and naturally lead to a crossed-product algebra of horizon observables.


Within holographic settings, the CPS framework has been used to recast gauge/gravity duality \cite{Parvizi:2025shq, Parvizi:2025wsg}. Extensions of the Iyer-Wald formalism to AdS/CFT subregion duality have yielded new perspectives on deriving Einstein’s equations from the entanglement entropy of the boundary CFT \cite{Faulkner:2013ica, Lashkari:2013koa, Faulkner:2017tkh, Haehl:2017sot}. These developments provide support for the Jafferis-Lewkowycz-Maldacena-Suh (JLMS) relation \cite{Jafferis:2015del} linking bulk and boundary relative entropies, clarifies how quantum information is encoded in gravitational charges \cite{Lashkari:2015hha}, and enable the formulation of bit thread constructions \cite{Agon:2020mvu, Das:2025fav}.

In the framework of Witten's open-string field theory (OSFT), analysing the CPS and its underlying symplectic structure is essential for understanding tachyon condensation. Recent work has successfully constructed the symplectic form in the CPS framework for OSFT on a ZZ-brane within $c=1$ string theory, a development that allowed precise calculation of the energy of the rolling tachyon solution \cite{Cho:2023khj}. Recent works by Bernardes, Erler, and F{\i}rat \cite{Bernardes:2025uzg, Bernardes:2025zzu, Bernardes:2025zkj, Bernardes:2026egp} (see also \cite{Ali:2026kmk}) proposed a covariant symplectic structure for generic \(L_\infty\)-Lagrangian field theories that is especially useful for nonlocal models, since it avoids the need for an explicit derivative expansion while reproducing the expected symplectic structure in standard examples.

\subsection{Motivation for the covariant phase space formalism}
\label{motdetails}

For the ordinary tensile open string, the target space noncommutativity is usually derived from the worldsheet operator formalism. In the presence of a constant Kalb-Ramond field, the open string boundary conditions become mixed, and the boundary two-point function acquires an antisymmetric part proportional to the Seiberg--Witten parameter \(\Theta^{\mu\nu}\). The equal-time endpoint commutator is then read off from this antisymmetric part of the boundary propagator \cite{Seiberg:1999vs}. We briefly review this standard derivation in Appendix~\ref{sec:open-string_propagator}. That derivation, however, relies crucially on structures special to the tensile theory: a nondegenerate worldsheet metric, the existence of an ordinary $2$-dim CFT, the holomorphic/antiholomorphic decomposition of the embedding fields, the usual oscillator expansion, and the logarithmic form of the propagator. In the tensionless limit \(T\to 0\), these ingredients cease to be available in their standard form. The worldsheet geometry degenerates to a Carrollian geometry, the left- and right-moving sectors no longer admit the usual separation, and the theory becomes ultralocal along the spatial direction of the worldsheet. As a result, the standard boundary-propagator argument for endpoint noncommutativity no longer works. 


The CPS viewpoint is particularly useful because it treats the tensile and tensionless theories within a single geometric language. In the tensile theory, it reproduces the usual Seiberg--Witten noncommutativity parameter from the boundary symplectic form. In the tensionless theory, where the free bulk symplectic structure becomes degenerate, it reveals that a constant antisymmetric background generates a symplectic form supported entirely on the boundary. The resulting endpoint noncommutativity is therefore not merely a deformation of the standard tensile result but the surviving remnant of the phase space itself in the Carrollian regime. When a boundary \(U(1)\) gauge field is present, the corresponding boundary contribution is incorporated most naturally using the CPS formalism with boundaries (CPSB).

\subsection{Summary of main results}
In this paper, we develop a CPS treatment of noncommutativity for open bosonic strings in constant antisymmetric backgrounds, with the aim of placing the usual tensile theory and the intrinsically tensionless theory within a single geometric framework. The central idea is that the noncommutative structure of the open string endpoints can be derived directly from the symplectic data on the space of classical solutions, without relying on the standard operator-based analysis of boundary correlators.

For the tensile open string in a constant Kalb-Ramond background, we show that the pre-symplectic current naturally splits into two qualitatively distinct pieces: a genuine bulk contribution coming from the kinetic term, and an exact contribution generated by the antisymmetric background field. After imposing the mixed open-string boundary conditions, the full symplectic structure reduces at the endpoints to a boundary form whose inverse reproduces the familiar Seiberg--Witten noncommutativity parameter,
\begin{equation}
\Theta^{\mu\nu}
= 2\pi\a'\left(\frac{1}{g+2\pi\a' \B}\right)^{\m\n}_A.
\end{equation}
or equivalently,
\begin{equation}
\mathbf{\Theta}
= -(2\pi\alpha')^2 (\bos{g}+2\pi\alpha' \bos{\B})^{-1} \bos{\B} (\bos{g}-2\pi\a' \bos{\B})^{-1}.
\end{equation}
Thus, in the tensile theory, the noncommutative geometry of the D-brane worldvolume emerges directly from the boundary symplectic structure of the CPS.
qWe then study the tensionless reduction of this result by taking the limit
\begin{equation}
T = \frac{1}{2\pi\alpha'} \to 0,
\qquad
\alpha' \to \infty.
\end{equation}
In this limit, the Seiberg--Witten parameter remains finite and approaches
\begin{equation}
\bos{\Theta}_{\text{tensionless}}
=
\lim_{\alpha' \to \infty}\bos{\Theta}
=
\bos{\B}^{-1},
\end{equation}
provided the antisymmetric background is invertible on the relevant brane directions. This provides an important bridge between the ordinary tensile description and the intrinsically tensionless theory.

The most significant result of the paper concerns the CPS of the intrinsically tensionless string itself. Starting from the ILST action, we show that in the absence of background antisymmetric fields the bulk pre-symplectic current vanishes identically, and therefore the corresponding symplectic form is degenerate
\begin{equation}
\bos{\Omega}_{0}=0.
\end{equation}
This implies that the free tensionless open string carries no nontrivial intrinsic bulk Poisson structure. In this sense, the ordinary propagating phase space collapses in the Carrollian regime.

Once a constant Kalb-Ramond background is included, however, the situation changes qualitatively. Although the free tensionless sector continues to contribute no nontrivial bulk symplectic form, the antisymmetric background generates an exact (pre)symplectic current, so that the full symplectic structure localises entirely on the boundary. The reduced boundary symplectic form takes the form
\begin{equation}
\bos{\Omega}_{\mathrm{bdry}} = \dfrac{1}{2}\B_{\mu\nu} \,\d X^\m \cw \d X^\nu ,
\end{equation}
and therefore the endpoint coordinates satisfy the noncommutative Poisson algebra
\begin{equation}
\{X^\mu, X^\nu\}
=
\B^{\mu\nu}.
\end{equation}
Equivalently, the intrinsic noncommutativity parameter of the tensionless theory is
\begin{equation}
\Theta^{\mu\nu}
=
\B^{\mu\nu}.
\end{equation}
This shows that, in the tensionless regime, noncommutativity is not merely a deformation of an otherwise nondegenerate bulk phase space. Rather, the boundary noncommutative algebra is the surviving physical remnant of the phase space itself.

We further extend the analysis by including a boundary $U(1)$ gauge field coupling for strings ending on a D$p$-brane. In this case, the appropriate framework is the CPSB. The resulting pre-symplectic current remains exact, and the boundary symplectic structure is controlled by the effective antisymmetric combination on the brane
\begin{equation}
\bos{\Omega}_{\text{bdry}}
= \frac{1}{2}\sf_{\mu\nu}\, \d X^\mu \cw \d X^\nu,
\end{equation}
where $\mathcal{F}_{\mu\nu}$ denotes the gauge invariant Born-Infeld type field strength induced on the boundary. The endpoint Poisson brackets are then governed by
\begin{equation}
\{X^\mu, X^\nu\} = \sf^{\mu\nu},
\qquad
\Theta^{\mu\nu} = \sf^{\mu\nu}.
\end{equation}
Thus, in the tensionless theory, the noncommutative geometry is determined more generally by the antisymmetric boundary data seen by the open-string endpoints.

\subsection*{Organisation of the paper.}

The paper is organised as follows. In Section \ref{sec:cps}, we provide a brief review of the Iyer-Wald CPS formalism and its extension to manifolds with boundaries. In Section \ref{sec:cps_tensilenew}, we apply this formalism to tensile strings in a constant Kalb-Ramond background, deriving the bulk and boundary symplectic structure, and recovering the Seiberg-Witten noncommutativity parameter. Section \ref{sec:cps_tensionless} contains the core analysis of tensionless strings, demonstrating the vanishing of the free bulk symplectic form, the emergence of purely boundary-supported noncommutativity in the presence of a constant $B$-field, and the subsequent effects of boundary gauge fields. We conclude in Section \ref{sec:cons} with a summary of our work and a discussion of future research directions.

\section{Covariant phase space formalism: a brief review}\label{sec:cps}

The canonical phase space  $\mathsf{P}$ of a theory defined in a global hyperbolic manifold $\M \cong \C \times \mathbb{R}$ is based on the set of initial data on a Cauchy surface $\C$ and the choice of Cauchy surface.  Hamiltonian mechanics is essential for analysing the dynamical behaviour of a theory, yet selecting a specific time direction breaks covariance and consequently destroys some of the symmetries of the underlying theory. To overcome this issue, it is useful to adopt a more geometric, mathematically structured approach, namely the covariant phase space (CPS) formalism \cite{Crnkovic:1986ex, Crnkovic:1987tz, Wald:1993nt, Iyer:1994ys, Zuckerman:1986vzu, Lee:1990nz}. It has deep roots in both physics and mathematics. The mathematical construction of the CPS formalism is based on the structural gr\'undlage of Anderson’s variational bicomplex \cite{Anderson1992IntroductionTT}. Here, we will discuss the formalism from the physics perspective, followed by a brief outline of the Iyer-Wald method \cite{Iyer:1994ys} and the one generalised by Harlow-Wu \cite{Harlow:2019yfa}.\\ 

Consider a $d$-dimensional spacetime manifold $\M$ on which the theory of some dynamical fields, collectively denoted by $\phi$ (these are tensors on $\M$ with suppressed indices), is defined. Let $\F$ be the functional space of all \emph{kinematically allowed} field configurations, called the \emph{field configuration space} or, in short, the \emph{field space}. The space $\F$ is an infinite-dimensional manifold, and each allowed field configuration is a point in $\F$.

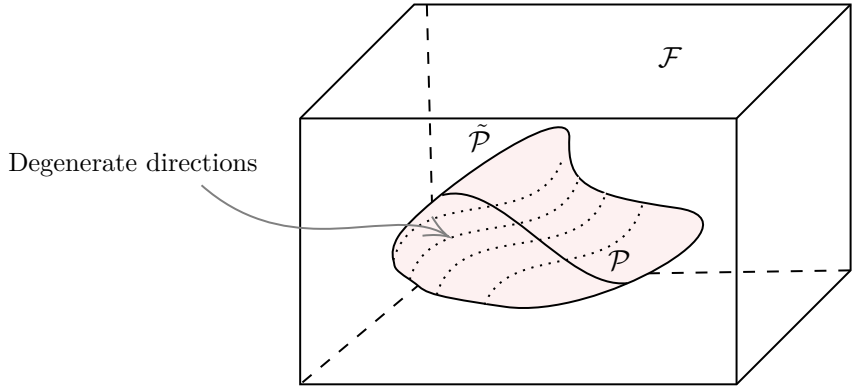
\begin{figure}[H]
\centering
\tikzset{every picture/.style={line width=0.75pt}} 

\begin{tikzpicture}[x=0.75pt,y=0.75pt,yscale=-1.1,xscale=1.1]
\path (600,225); 

\draw   (215.14,59.28) -- (267.43,7) -- (467.5,7) -- (467.5,128.99) -- (415.22,181.28) -- (215.14,181.28) -- cycle ; \draw   (467.5,7) -- (415.22,59.28) -- (215.14,59.28) ; \draw   (415.22,59.28) -- (415.22,181.28) ;
\draw  [dash pattern={on 4.5pt off 4.5pt}]  (273.22,7) -- (276.44,131.81) ;
\draw  [dash pattern={on 4.5pt off 4.5pt}]  (276.44,131.81) -- (467.5,128.99) ;
\draw  [dash pattern={on 4.5pt off 4.5pt}]  (270.52,134.85) -- (215.14,181.28) ;
\draw  [fill={rgb, 255:red, 253; green, 241; blue, 241 }  ,fill opacity=1 ] (260,113.61) .. controls (269.24,98.17) and (339.56,47.46) .. (338.54,68.25) .. controls (337.52,89.03) and (345.73,94.39) .. (387.83,100.37) .. controls (429.92,106.35) and (351.39,151.71) .. (309.29,145.73) .. controls (267.2,139.75) and (273.61,138.56) .. (265.4,133.21) .. controls (257.18,127.86) and (258.79,127.56) .. (257.89,124.29) .. controls (256.99,121.03) and (257.7,117.46) .. (260,113.61) -- cycle ;
\draw [fill={rgb, 255:red, 253; green, 241; blue, 241 }  ,fill opacity=1 ] [dash pattern={on 0.84pt off 2.51pt}]  (277.77,140.2) .. controls (288.03,103.98) and (345.3,128.97) .. (355.56,92.75) ;
\draw [fill={rgb, 255:red, 253; green, 241; blue, 241 }  ,fill opacity=1 ] [dash pattern={on 0.84pt off 2.51pt}]  (299.75,144.77) .. controls (311.1,115.07) and (361.71,135.31) .. (371.97,99.09) ;
\draw [fill={rgb, 255:red, 253; green, 241; blue, 241 }  ,fill opacity=1 ] [dash pattern={on 0.84pt off 2.51pt}]  (265.4,133.21) .. controls (275.66,96.99) and (332.93,121.98) .. (343.19,85.76) ;
\draw [fill={rgb, 255:red, 253; green, 241; blue, 241 }  ,fill opacity=1 ] [dash pattern={on 0.84pt off 2.51pt}]  (257.89,124.29) .. controls (268.15,88.07) and (325.42,113.06) .. (335.68,76.84) ;
\draw    (280.24,94.52) .. controls (306.87,86.91) and (337.31,137.9) .. (365.21,134.85) ;
\draw [color={rgb, 255:red, 128; green, 128; blue, 128 }  ,draw opacity=1 ]   (169.91,89.95) .. controls (188.21,106.1) and (207.04,109.78) .. (224.49,109.94) .. controls (247.34,110.15) and (267.85,104.34) .. (281.72,112.59) ;
\draw [shift={(283.2,113.55)}, rotate = 206.3] [color={rgb, 255:red, 128; green, 128; blue, 128 }  ,draw opacity=1 ][line width=0.75]    (10.93,-4.9) .. controls (6.95,-2.3) and (3.31,-0.67) .. (0,0) .. controls (3.31,0.67) and (6.95,2.3) .. (10.93,4.9)   ;

\draw (377.99,25.4) node [anchor=north west][inner sep=0.75pt]   [align=left] {$\displaystyle \mathcal{F}$};
\draw (290.99,61.05) node [anchor=north west][inner sep=0.75pt]   [align=left] {$\displaystyle \tilde{\mathcal{P}}$};
\draw (80,73.82) node[anchor=north west, inner sep=0.75pt, font=\small, align=left] {\small \textnormal{Degenerate directions}};
\draw (355.02,118.48) node [anchor=north west][inner sep=0.75pt]   [align=left] {$\displaystyle \mathcal{P}$};

\end{tikzpicture}

\caption{\textit{A schematic diagram depicting the symplectic reduction. The pre-phase space $\til{\P}\subset\F$ is shown in beige, consisting of the solutions of the e.o.m. The associated pre-symplectic form $\til{\bos{\Omega}}$ has degenerated directions shown with dotted curves generated by the zero modes $Z$. By symplectic reduction, one obtains the phase space $\P$ shown with a solid curve.}}
\label{fig:phase-space-reduction}
\end{figure}

The subspace (or submanifold) of the field space consisting of field configurations that satisfy the equations of motion is denoted by $\til{\P} \subset \F$ and is called the \emph{pre-phase space}\footnote{$\til{\P}$ is also referred to as the \emph{solution space} \cite{Hajian:2015xlp, Margalef-Bentabol:2020teu}.} (see Figure \ref{fig:phase-space-reduction}). So, pre-phase space is the functional space of all the \emph{dynamically} allowed field configurations. One can construct the \emph{pre-symplectic} form $\til{\bos{\Omega}}$ on $\til{\P}$, which is closed by definition but may be degenerate in nature, which is the reason for adding `pre' before the symplectic form. This definition of (pre)-symplectic form does not pick out a preferred time direction or coordinates and is hence covariant. It can be shown that the symplectic form obtained from CPS formalism is equivalent to the standard (canonical) symplectic form defined on the cotangent bundle $\T^*\F$ (i.e., $\mathsf{P}$). For a mathematical proof of the equivalence between the covariant and canonical symplectic forms, see \cite{Margalef-Bentabol:2022zso}. In Appendix \ref{sec:can_vs_cps}, we provide a brief summary of the comparison between the canonical phase space structure and the CPS structure for the sake of completeness.

 The fact that the pre-symplectic form is degenerate arises from gauge redundancy (continuous local symmetries): multiple solutions may correspond to the same physical state. A vector $Z$ on $\til{\P}$ is called a \emph{zero mode} of $\til{\bos{\Omega}}$, if it connects two such solutions. Mathematically speaking, $Z.\til{\bos{\Omega}} = 0$. These directions are often regarded as degenerate directions (the dotted curves shown in Figure \ref{fig:phase-space-reduction}).  All such zero modes of $\til{\bos{\Omega}}$ form a Lie algebra that generates a group ($\mathscr{G}$) of gauge symmetries \cite{Harlow:2019yfa}. In the presence of gauge symmetries, one needs to take the quotient of $\til{\P}$ by the symmetry group $\mathscr{G}$ to obtain the `\emph{covariant phase space}' $\P$ ($\equiv \til{\P}/\mathscr{G}$), which actually denotes the set of equivalence classes of $\til{\P}$. One can define a (projection) map $\pi: \til{\P} \rightarrow \P$ that assigns each element of  $\til{\P}$ to its equivalence class. It also helps define a non-degenerate closed $2$-form $\bos{\Omega}$ on $\P$,\footnote{$\bos{\Omega}$ is related to the pre-symplectic form via the pullback of the projection map $\pi$ as follows \cite{Reyes:2004zz}
 \be 
 \pi^*\bos{\Omega} = \til{\bos{\Omega}}. \nn
 \ee} making $\P$ a symplectic manifold, the (covariant) phase space. In symplectic geometry, this procedure is called the \emph{symplectic reduction}.\footnote{Here we are quotientising the gauge group after we restrict ourselves to the solution space $\til{\P}$ of the whole field space $\F$ following \cite{Zuckerman:1986vzu, Crnkovic:1986ex, Crnkovic:1987tz, Harlow:2019yfa}. Instead, one can do the symplectic reduction before restricting to the dynamically allowed subspace of $\F$ \cite{Lee:1990nz}.} In the absence of gauge symmetries, $\til{\P}$ and $\P$ eventually coincide.

In the following, we introduce the notations and conventions that are used throughout this paper.

 \paragraph{Notation and conventions.} 
 \begin{itemize}
\item Any bold font Latin/Greek letter (e.g. $\bos{K}, \bos{\a}$) denotes a form either in spacetime manifold $\M$ or field spcae $\F$. 

\item The spacetime field variations $\d \phi$ construct the basis of the tangent bundle $\T\F$. 

\item In general, a form $\bos{K}$ is a bigraded form of degree $(p;q)$, which means $\bos{K}$ is a $p$-form in spacetime and $q$-form in the field space. Sometimes, we may use the convention $p$-form, which refers to the spacetime (or worldsheet in our case) unless specified.
 
\item In the CPS formalism, one has exterior calculus on both $\M$ and $\F$.\footnote{The field space is assumed to have sufficient structures to be an infinite dimensional Banach manifold such that the exterior calculus can be defined.} The necessary operators we will need are the exterior derivatives:
\begin{enumerate}[i.]
    \item Spacetime exterior derivative is denoted by $\ed$;

    \item Field space exterior derivative is denoted by $\d$, which corresponds to the variation on the spacetime.
\end{enumerate}
 Both of these exterior derivatives are nilpotent, i.e., $\ed^2 =0 = \d^2$. In contrast to the mathematics literature, where these derivatives are considered to anticommute, we will assume they commute:\footnote{In most of the mathematics literature (e.g. \cite{Anderson1992IntroductionTT, Barnich:2001jy, Barnich:2007bf}), the total differential is defined as $\ed+\d$, which leads to the anicommutation of the derivatives. Instead, we will consider the total differential to be $\ed + (-1)^{p+q} \d$ so that $\ed$ and $\d$ commute. It simplifies our calculation. In general, the bi-gradded commutator between a $(p;q)$-form $\bos{\a}$ and a $(r;s)$-form $\bos{\b}$ is given by \cite{Speziale:2025lkm},
 \be
 [\bos{\a},\bos{\b}] = \bos{\a \b} - (-1)^{pr+qs} \bos{\b\a}\,. \nn
 \ee
}
\be
 [\ed,\d]=0.
\ee

\item We denote the spacetime wedge product by $\w$ and the field space wedge product by $\cw$.

 A $(p,q)$-form $\bos{\a}$ in spacetime is expressed as
 \begin{subequations}
 \be
 \bos{\a}(x) = \dfrac{1}{p!}\, \a_{\m_1 \m_2 \cdots \m_p}(x)\, \ed x^{\m_1} \w \ed x^{\m_2} \cdots \w \ed x^{\m_p},
 \ee
 and in the field space is expressed as
 \be
 \bos{\a}(\phi) = \dfrac{1}{q!}\,  \a_{I_1 I_2 \cdots I_q}(\phi)\, \d \phi^{I_1} \cw \d \phi^{I_2} \cdots \cw \d \phi^{I_q},
 \ee
 \end{subequations}
 where $\phi^I$ denotes the field components, and the dynamical fields are given by $\phi=\{\phi^I\}$.
 \end{itemize}



\subsection{Iyer-Wald formalism}

  The construction is as follows. Given a general theory on a $d$-dimensional manifold $\M$, one begins with the action 
\begin{equation}
	S[\phi] = \int_{\M} \bos{L}(\phi,\psi),
\end{equation}
where $\bos{L}$ is the Lagrangian $d$-form, $\phi$ denotes all the dynamical fields and $\psi$ denotes some background fields which do not play any role in the dynamics of the theory. Varying the action leads to
\begin{equation}
	\delta \bos{L}[\phi] = \bos{E}_{\phi}  \delta \phi + \mathrm{d} \bost_{\rm{IW}} \left( \delta \phi, \phi \right),
\end{equation}
where $\delta \phi$ is a generic field perturbation in spacetime. The equations of motion follow from $\boldsymbol{E}_{\phi} = 0$. 
The quantity $\boldsymbol{\theta}_{\mathrm{IW}}$, referred to as the symplectic potential, 
is a $(d-1)$-form on spacetime $\mathcal{M}$ and simultaneously a one-form on the phase space, 
i.e., a $(d-1;1)$-form.\footnote{The symplectic potential is not unique and is ambiguous up to two terms \cite{Iyer:1994ys}
\be
 \bost_{IW} \rightarrow \bost_{IW} + \d \bos{W} + \ed \bos{Y} \nn,
\ee
where $\bos{W}$ is a $(d-1;2)$-form and $\bos{Y}$ is a $(d-2;1)$-form.} Then, one can define the quantity called symplectic current as
\begin{equation}\label{sympcurrent}
	\bom_{\rm{IW}}(\delta_{1} \phi, \delta_{2} \phi,  \phi) = \delta_{1} \bost_{\rm{IW}}(\delta_{2} \phi, \phi) - \delta_{2} \bost_{\rm{IW}} (\delta_{1} \phi,  \phi).
\end{equation}
Here, $\delta_{1} \phi$ and $\delta_{2} \phi$ are two arbitrary field perturbations. It is evident that the symplectic current $\bom_{\rm{IW}}$ is a $(d-1;2)$-form. Having found this current, the next step is quite simple: one defines the symplectic form\footnote{To be more precise, one gets the pre-symplectic form as mentioned earlier in this section. For this part, we will follow \cite{Iyer:1994ys} to call it the symplectic form by assuming that the symplectic reduction has already been done.} by integrating the symplectic current over a codimension-one surface $\C$,
\begin{equation}
	\bos{\Omega}_{\rm{IW}} (\d_{1} \phi, \d_{2} \phi,  \phi) = \int_{\C} \bom_{\rm{IW}}(\d_{1} \phi, \d_{2} \phi,  \phi),
\end{equation}
which is closed and nondegenerate. If the codim-1 surface is a Cauchy surface, the symplectic form does not depend on it. This yields a well defined mathematical structure that relates to how the Hamiltonian varies in the following way
\be
 \bos{\Omega}_{\rm{IW}} (\d_{1} \phi, \d_{2} \phi,  \phi) = \d \H\,.
\ee
The Noether current, which is a $(d-1)$-form, associated with a symmetry generator $\xi^\mu$, is defined as
\begin{equation}\label{noethercurrent}
	\bos{J}_{\xi} \equiv \bost_{\rm{IW}} (\L_\xi \phi, \phi) - \xi .\bos{L} ~ .
\end{equation}
As the Noether current is conserved (i.e., closed: $\ed \bos{J_\xi=0}$), it can be locally expressed as the exterior derivative of another $(d-2)$-form, the Noether charge 
\begin{equation}\label{noethercharge}
	\bos{J}_{\xi} = \mathrm{d} \bos{ Q}_{\xi}.
\end{equation}
In diffeomorphism-invariant theories, covariance implies that 
$\delta_\xi \phi = \mathcal{L}_\xi \phi$, where $\xi^\mu$ generates 
diffeomorphisms. Accordingly, the symplectic current can be expressed as
\begin{equation}\label{sympcurrent_diffs}
	\bom_{\rm{IW}}(\delta \phi, \mathcal{L}_\xi \phi,  \phi) = \delta \bost_{\rm{IW}}(\mathcal{L}_\xi\phi,\phi)- \mathcal{L}_\xi \bost_{\rm{IW}} (\delta \phi,  \phi).
\end{equation}
It can also be expressed in terms of the Noether current and Noether charge as follows:
\begin{align}\label{sympcurrent_diffs_2}
	\bom_{\rm{IW}} &= \delta \boldsymbol{\mathrm{J}}_{\xi} - \ed (\xi.\bost_{\rm{IW}}) = \delta (\ed\bos{ \mathrm{Q}}_{\xi}) -  \ed (\xi.\bost_{\rm{IW}}), \nonumber \\
	&= \ed (\delta \bos{ \mathrm{Q}}_{\xi} - \xi.\bost_{\rm{IW}}).
\end{align}
In the last line, we have used $[\ed,\d]=0$. Now, the change in Hamiltonian is given by
\begin{equation}\label{hamiltonianIW}
	\delta \H_\xi = \int_{\pa\C} (\delta \boldsymbol{ \mathrm{Q}}_{\xi} - \xi.\bost_{\rm{IW}}).
\end{equation}

\subsection{CPS with boundaries}
\begin{figure}[H]
\centering

\tikzset{every picture/.style={line width=0.75pt}} 

\begin{tikzpicture}[x=0.75pt,y=0.75pt,yscale=-1,xscale=1]

\draw    (404.21,83.5) .. controls (393.99,63.77) and (367.33,77.44) .. (318.42,65.47) .. controls (269.5,53.5) and (241.38,59.82) .. (238.16,68.4) ;
\draw    (404.21,83.5) .. controls (376.29,95.1) and (312.03,83.69) .. (293.84,71.66) .. controls (275.64,59.63) and (253.83,76.3) .. (238.16,68.4) ;
\draw    (231.03,221.6) .. controls (248.64,232) and (292.51,215.64) .. (319.74,227.55) .. controls (346.97,239.46) and (392.79,220.14) .. (412.53,211.89) ;
\draw [color={rgb, 255:red, 74; green, 74; blue, 74 }  ,draw opacity=1 ]   (412.53,211.89) .. controls (390.17,199.66) and (384.1,219.23) .. (349.53,216.86) .. controls (314.95,214.5) and (246.11,206.16) .. (231.03,221.6) ;
\draw    (238.16,68.4) .. controls (264.2,130.79) and (220.66,167.94) .. (231.03,221.6) ;
\draw    (404.21,83.5) .. controls (408.48,153.78) and (369.2,159.09) .. (412.53,211.89) ;
\draw [color={rgb, 255:red, 74; green, 144; blue, 226 }  ,draw opacity=1 ][fill={rgb, 255:red, 252; green, 241; blue, 240 }  ,fill opacity=1 ]   (243.28,141.4) .. controls (240.03,133.11) and (291.77,132.41) .. (335.68,135.57) .. controls (366.67,137.8) and (393.76,141.95) .. (394.82,146.71) ;
\draw [color={rgb, 255:red, 74; green, 144; blue, 226 }  ,draw opacity=1 ][fill={rgb, 255:red, 252; green, 241; blue, 240 }  ,fill opacity=1 ]   (394.82,146.71) .. controls (397.14,156.32) and (362.03,155.73) .. (335.27,154.81) .. controls (322.63,154.38) and (311.85,153.87) .. (307.74,154.33) .. controls (294.94,155.77) and (244.56,160.86) .. (243.28,141.4) ;
\draw  [dash pattern={on 4.5pt off 4.5pt}]  (231.76,22) .. controls (236.88,35.27) and (231.76,54.72) .. (238.16,68.4) ;
\draw  [dash pattern={on 4.5pt off 4.5pt}]  (397.8,37.1) .. controls (402.93,50.36) and (397.8,69.82) .. (404.21,83.5) ;
\draw  [dash pattern={on 4.5pt off 4.5pt}]  (231.03,221.6) .. controls (236.16,234.86) and (231.03,254.32) .. (237.44,268) ;
\draw  [dash pattern={on 4.5pt off 4.5pt}]  (412.53,211.89) .. controls (417.87,228.96) and (420.43,246.65) .. (418.93,258.29) ;
\draw [color={rgb, 255:red, 128; green, 128; blue, 128 }  ,draw opacity=1 ]   (201.88,130.79) .. controls (223.19,111.26) and (257.08,120.26) .. (268.83,129.62) ;
\draw [shift={(270.18,130.79)}, rotate = 223.52] [color={rgb, 255:red, 128; green, 128; blue, 128 }  ,draw opacity=1 ][line width=0.75]    (10.93,-4.9) .. controls (6.95,-2.3) and (3.31,-0.67) .. (0,0) .. controls (3.31,0.67) and (6.95,2.3) .. (10.93,4.9)   ;

\draw (317.58,43.47) node [anchor=north west][inner sep=0.75pt]    {$\mathcal{C}_{+}$};
\draw (308.26,235.4) node [anchor=north west][inner sep=0.75pt]    {$\mathcal{C}_{-}$};
\draw (300.65,136.34) node [anchor=north west][inner sep=0.75pt]    {$\mathcal{C}$};
\draw (408.02,111.69) node [anchor=north west][inner sep=0.75pt]    {$\Gamma $};
\draw (188.74,131.15) node [anchor=north west][inner sep=0.75pt]    {$\partial \mathcal{C}$};
\draw (351.65,174.34) node [anchor=north west][inner sep=0.75pt]    {$\mathcal{M}$};

\end{tikzpicture}

\caption{\textit{A portion of the manifold $\M$ with boundary $\pa\M = \Gamma \cup \C_{-} \cup \C_{+}$. The surfaces $\C_{+}$ and $\C_{-}$ denote the future and past temporal boundaries, respectively, while $\Gamma$ denotes the spatial boundary. The beige patch represents a Cauchy surface $\C$, and its boundary $\pa\C$ is highlighted in blue.}}
\label{bndy}
\end{figure}
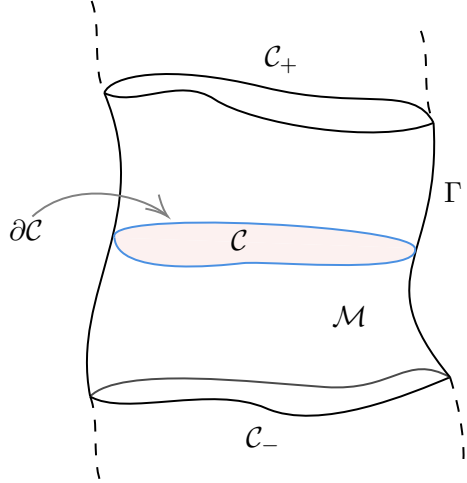

The CPS formalism has been generalised by incorporating a boundary Lagrangian $\ell$, resulting in the development of the \textit{covariant phase space formalism with boundaries} (CPSB) \cite{Harlow:2019yfa}.\footnote{A mathematically rigorous formulation of this formalism can be found in \cite{Margalef-Bentabol:2020teu}.} Similar to the IW method, here also we consider a theory defined on a $d$-dimensional manifold $\M$ with boundary $\pa\M$. In this framework, the action is expressed as 
\begin{equation}\label{actioncpsb}
	S[\phi] = \int_{\mathcal{M}} \bos{L} \big(\phi, \psi \big) + \int_{\partial \mathcal{M}} \bos{\ell} \big(\phi, \psi \big),
\end{equation}
where $\bos{L}$ is the bulk Lagrangian $d$-form, $\bos{\ell}$ is the boundary Lagrangian $(d-1)$-form, $\phi$ denotes the dynamical field collectively and $\psi$ is the background field. The formulation imposes that the action is well-posed; specifically, $\phi$ satisfies the equations of motion (e.o.m.) if and only if the action $S[\phi]$ is stationary, accounting for some boundary terms at future and past slices. The variation of the bulk Lagrangian can be expressed as 
\begin{equation}
	\d \bos{L} = \bos{E}_{\phi} \d \phi + \ed  \bost_{\rm{HW}},
\end{equation}
where $\bost_{\rm{HW}}$ is the symplectic potential. Thus, the variation of the action becomes 
\begin{equation}
	\d S = \int_{\M} \bos{E}_{\phi} \d \phi + \int_{\C_{\pm}} \big( \bost_{\rm{HW}} + \d \bos{\ell} \big) + \int_{\Gamma} \big( \bost_{\rm{HW}} + \d \bos{\ell} \big).
\end{equation}
The boundary of the manifold $\M$ is expressed as $\pa\M= \Gamma \cup \C_{-} \cup \C_{+}$, where $\C_{\pm}$ denote the future and past boundaries, respectively, and $\Gamma$ is the spatial boundary (see Figure \ref{bndy}).

The stationarity condition of \( S[\phi] \) implies that \( \d S[\phi] = 0 \), which leads to two important equations: the first is the equation of motion (e.o.m.), given by
\[
\bos{E}_\phi = 0  
\]
The second equation is nontrivial and represents a new aspect compared to the IW method,
\be
\big( \bost_{\rm{HW}} + \delta \boldsymbol{\ell} \big)\mid_{\Gamma} = \mathrm{d} \bos{\sc}
\ee
for any \((d-2)\)-form \(\bos{\mathsf{C}}\), which is also referred to as the \emph{corner} term. In the IW method, \(\bos{\mathsf{C}}\) is set to zero, which is a special case of CPSB. Next, one can define the pre-symplectic current as follows,
\be\label{eq:symp_current_HW}
\bom_{\rm{HW}} \equiv \d(\bost_{\rm{HW}} - \ed \bos{\sc})\mid_{\tilde{\P}} 
\ee
By definition, this current is biclosed, i.e., $\ed\bom=0$ and $ \d\bom=0$. Integrating the pre-symplectic current over a Cauchy slice yields the pre-symplectic form,
\be
\tilde{\mathbf{\Omega}}_{\rm{HW}} = \int_{\C} \bom_{\rm{HW}}
\ee
This form is closed and independent of the slice \(\C\). The pre-symplectic form is utilised to construct the Hamiltonian associated with this CPSB. 

Consider a diffeomorphism generator \(\xi^{\mu}\) that preserves the boundary conditions and where the Lagrangians are covariant with respect to \(\xi^{\mu}\). The Hamiltonian \(\H_{\xi}\) is defined by the relation,\footnote{Finding an $\H_\xi$ obeying this relation is sufficient to define the Hamiltonian on $\P$ as $\til{\bos{\Omega}}$ uniquely constructs $\bos{\Omega}$ on $\P$ via the projection map $\pi$ and for any zero mode $Z$ of $\til{\bos{\Omega}}$ one has
\[
Z.\delta \H_{\xi} = \til{\bos{\Omega}}(Z,X_\xi) = 0,
\]
which ensures that $\H_\xi$ will be a well defined function on $\P$.}
\be\label{hamiltonian}
\delta \H_{\xi} = - X_{\xi}.\tilde{\mathbf{\Omega}}_{\rm{HW}},
\ee
where the vector field in the field space \(\F\) is expressed as,
\[
X_{\xi} \equiv \int d^{d}\mathrm{x} \; \d_{\xi}\phi^{I}(x) \dfrac{\delta}{\delta \phi^{I}(x)}.
\]
With the help of the field space vector field, one can define the Noether current $(d-1)$-form as
\be
\boldsymbol{\mathrm{J}}_{\xi} \equiv X_{\xi}.\bost_{\rm{HW}} - \xi. \boldsymbol{ \mathrm{L}}.
\ee
A straightforward calculation using the previous three equations yields,
\be\label{xom}
-X_{\xi}.\tilde{\mathbf{\Omega}}_{\rm{HW}} = \d \left(\int_{\C} \bos{J}_{\xi} + \int_{\pa\C} \big(\xi.\bos{\ell} - X_{\xi}.\bos{ \sc}\big) \right).
\ee
From equations \(\eqref{hamiltonian}\) and \(\eqref{xom}\), we can derive the Hamiltonian:
\be
\H_{\xi} = \int_{\C} \bos{J}_{\xi} + \int_{\pa\C} \big(\xi.\bos{\ell} - X_{\xi}.\bos{ \sc}\big).
\ee
This result holds for any higher derivative theory. Some examples can be found in \cite{Harlow:2019yfa}. 

In certain special theories, such as general relativity, \(\bos{L}\) is covariant under arbitrary \(\xi^{\mu}\). In such cases, we have \(\ed\bos{J}_{\xi} = 0\), and \(\bos{J}_{\xi}\) becomes an exact form. Therefore, we can define a \((d-2)\)-form \(\bos{Q}_{\xi}\) such that,
\be
\bos{J}_{\xi} = \ed \bos{Q}_{\xi},
\ee
with \(\bos{Q}_{\xi}\) representing the Noether charge form. For these special theories, the Hamiltonian simplifies to,
\be
\H_{\xi} = \int_{\pa \C} \left( \bos{Q}_{\xi} + \xi.\bos{\ell} - X_{\xi}.\bos{ \sc} \right),
\ee
indicating that the Hamiltonian is totally a boundary term.


\paragraph{Gist of the two formalisms.}

\paragraph{Iyer-Wald formalism.}
Standard Hamiltonian mechanics picks a preferred time and breaks covariance. The Iyer-Wald (IW) formalism keeps covariance by taking the \textit{space of all solutions to the equations of motion} as the phase space. It gives a geometric derivation of conserved charges and Hamiltonians directly from the Lagrangian.

\begin{itemize}
    \item \textit{From fields to phase space.} Start with the space of fields $\mathcal{F}$ and impose the equations of motion $\mathbf{E}_{\phi}=0$ to obtain the CPS $\mathcal{P}$.
    \item \textit{The symplectic structure.} Varying the Lagrangian defines the symplectic potential $\bost_{\text{IW}}$ and hence the symplectic current $\bom_{\text{IW}}$, whose integral over a Cauchy surface $\C$ gives the symplectic form $\bos{\Omega}_{\text{IW}}$, the basic geometric structure on phase space.
    \item \textit{Symmetry and charges.} In diffeomorphism-invariant theories, every generator $\xi$ determines a Noether current $\bos{J}_{\xi}$ and Noether charge $\bos{Q}_{\xi}$, allowing the Hamiltonian variation to be written entirely in terms of these.
\end{itemize}

\paragraph{CPS with boundaries (CPSB).}
In systems with boundaries, especially in holography, the original IW formalism becomes ambiguous. Harlow and Wu (HW) extended it by adding a boundary term $\boldsymbol{\ell}$ to the action to ensure a well-posed variational principle. 

\begin{itemize}
    \item \textit{Stationarity with boundaries.} The full action has a boundary term $\int_{\partial \mathcal{M}} \bos{\ell}$. A well-posed variational principle requires a boundary consistency condition relating the symplectic potential to a $(d-2)$-form $\bos{\mathsf{C}}$.
    \item \textit{The refined Hamiltonian.} The Hamiltonian is adjusted to include boundary contributions from $\bos{\ell}$ and $\bos{\mathsf{C}}$, so that it still correctly generates symmetries in the presence of edges.
    \item \textit{Gravity as a boundary term.} For diffeomorphism covariant theories like GR, the bulk Noether current is exact ($\bos{J}_{\xi} = \ed\bos{Q}_{\xi}$), so the Hamiltonian becomes an integral over the boundary of the Cauchy surface $\partial \C$, supporting the holographic principle.
\end{itemize}
\section{CPS for tensile strings in the presence of a uniform Kalb--Ramond field}\label{sec:cps_tensilenew}

In this section, we apply the CPS machinery to study the specific physical system: a tensile string propagating through a target spacetime with a constant Kalb-Ramond background. To this end, our analysis naturally begins with the standard Polyakov action
\begin{equation}
\begin{split}
S_{\text{tensile}} 
&= - \frac{T}{2} \int_\Sigma d^2\sp\left( 
\sqrt{-g}\, g^{ab} \,\pa_a X^\m \pa_b X^\n \,g_{\m\n} 
+ \veps^{ab} B_{\m\n} \,\pa_a X^\m \pa_b X^\n
\right). \nn
\end{split}
\end{equation}
In the conformal gauge, the action simplifies to
\be
 S_{\text{conf}} = - \frac{T}{2} \int_\Sigma d^2\sp\left( 
\eta^{ab} \,\pa_a X^\m \pa_b X^\n \,g_{\m\n} 
+ \veps^{ab} B_{\m\n} \,\pa_a X^\m \pa_b X^\n 
\right),
\ee
where $\eta^{ab}$ represents the inverse of the flat worldsheet metric characteristic of the conformal gauge. From this simplified action, we can easily read off the corresponding Lagrangian $2$-form
\be
 \bos{L} = - \frac{T}{2} \left( \eta^{ab} \,\pa_a X^\m \pa_b X^\n \,g_{\m\n} 
+ \veps^{ab} B_{\m\n} \,\pa_a X^\m \pa_b X^\n  \right).
\ee
The core of the CPS method involves taking the first variation of the Lagrangian form. This variation naturally splits into two distinct parts: a bulk term that dictates the equations of motion (e.o.m.), and an exact boundary term that defines the symplectic potential. Mathematically, this is expressed as
\be
 \d \bos{L} = \bos{E}_\m\,\d X^\m + \ed \bost(X,\d X). \nn
\ee
By isolating the bulk piece, we recover the standard wave equation for a free string, as the constant $B$-field topological term does not contribute to the bulk dynamics
\be
 \bos{E}_\m = T \eta^{ab} g_{\m\n} \pa_a\pa_b X^\n =0.
\ee
Meanwhile, the boundary piece yields the symplectic potential $\theta^b$, which encodes the foundational phase space structure before taking the second variation
\be
 \theta^b = -T \left( \eta^{ab} g_{\m\n} + \veps^{ab}B_{\m\n} \right)\pa_a X^\m \d X^\n.
\ee
To construct the symplectic current $\omega^b$, we apply an exterior derivative in the configuration space of the fields. In practice, this means taking two independent, anti-symmetrised variations ($\delta_1$ and $\delta_2$) of the symplectic potential. We calculate
\begin{align*}
\omega^b(X; \delta_1 X, \delta_2 X) &= \delta_1 \theta^b(X, \delta_2 X) - \delta_2 \theta^b(X, \delta_1 X) \\
  &= -T \left( \eta^{ab} g_{\m\n} + \veps^{ab}B_{\m\n} \right)\big( \pa_a \d_1 X^\m \d_2 X^\n - \pa_a \d_2 X^\m \d_1 X^\n \big)\\
  &= T \left( \eta^{ab} g_{\m\n} + \veps^{ab}B_{\m\n} \right)\big( \pa_a \d_2 X^\m \d_1 X^\n - \pa_a \d_1 X^\m \d_2 X^\n \big) \\
  &=T \left( \eta^{ab} g_{\m\n} + \veps^{ab}B_{\m\n} \right) \pa_a \d X^\m \cw \d X^\n .
\end{align*}
At this stage, it is highly insightful to dissect the symplectic current into contributions arising from the spacetime metric $g$ and the antisymmetric $B$-field. Let's look closely at the $B$-field contribution, $\omega^b_B$. Because of the antisymmetry of the field, we can cleverly rewrite it as an exact total worldsheet derivative (extracting a factor of $1/2$)
\be
\omega^b_B=T \veps^{ab} B_{\m\n}\pa_a \d X^\m \cw \d X^\n = \frac{T}{2} \veps^{ab} B_{\m\n} \pa_a (\d X^\m \cw \d X^\n ).
\ee
Notice that this $B$-field component exactly gives the boundary-only result. The tensile string gives a non-vanishing bulk contribution generated by the spacetime metric
\be
\omega^b_g = T \eta^{ab}\eta_{\m\n} \,\pa_a\d X^\m \cw \d X^\n. 
\ee
The total symplectic current $\omega^b$ can be cleanly organised as the sum of a bulk current and the derivative of a boundary term
\begin{subequations}
\be
 \omega^b = \omega^b_g + \pa_a k_B^{ab} 
\ee
with
\be
 k_B^{ab} =  \frac{T}{2}  \veps^{ab}B_{\m\n} \d X^\m \cw \d X^\n .
\ee
\end{subequations}
In literature, the quantity $\bos{k}$ is called the \emph{surface charge} form.
Integrating the symplectic current over a Cauchy slice $\C$ on the worldsheet yields the fundamental symplectic form $\bos{\Omega}$ of the theory. The tensile string's symplectic form possesses both bulk and boundary characteristics
\begin{equation}
\begin{split}
\bos{\Omega}&= \int_\C \omega_g+ \int_{ \mathcal{C}} \omega_B \\
&=\int_\C \omega_g +  \Big[k_B\Big]_{\partial C}\\
&=\int_\C  T \eta^{ab}\eta_{\m\n} \,\pa_a\d X^\m \cw \d X^\n + \Bigg[\frac{T}{2} B_{\m\n} \,\d X^\m \cw \d X^\n \Bigg]_{\partial C}.
\end{split}
\end{equation}
\subsection{Noncommutativity from the symplectic structure of open strings}
In this section, we transition our focus to open strings. A fascinating consequence of turning on a constant background $B$-field is the emergence of noncommutative spacetime geometry localised at the endpoints of an open string. We will now rigorously derive this noncommutative parameter, $\Theta^{\mu\nu}$, directly from the underlying symplectic architecture we just explored. 
\paragraph{Action and boundary conditions.}
We resume with the Polyakov action in the conformal gauge, explicitly parameterised with the inverse string tension $2\pi\alpha'$
\begin{equation}
S_\text{tensile}=-\frac{1}{4\pi\alpha'}\int d\tau d\sigma
\left(-
g_{\mu\nu}(\dot X^\mu\dot X^\nu-X'^\mu X'^\nu)
+\veps^{ab} B_{\m\n} \,\pa_a X^\m \pa_b X^\n
\right).
\end{equation}
As we are going to compare the result with \cite{Seiberg:1999vs}, it is better to use the same action. We can do that by rescaling the Kalb-Ramond field as follows,
\be\label{eq:rescale_B_tensile}
\B_{\m\n} = -\dfrac{1}{2\pi\a'} B_{\m\n},
\ee
and rewriting the action in terms of the rescaled field as
\begin{equation}
S_\text{tensile}=\frac{1}{4\pi\alpha'}\int d\tau d\sigma
\left(
g_{\mu\nu}(\dot X^\mu\dot X^\nu-X'^\mu X'^\nu)
+ 2\pi\a' \veps^{ab} \B_{\m\n} \,\pa_a X^\m \pa_b X^\n
\right).
\end{equation}
By varying the action with respect to the worldsheet time derivative ($\dot {X} ^\mu$), we obtain the generalised canonical momentum density $\Pi_\mu$. The presence of the $B$-field profoundly modifies the standard momentum definition by mixing in spatial worldsheet derivatives
\begin{equation}\label{eq:tensilemomentum}
\Pi_\mu=\frac{\partial L}{\partial \dot X^\mu}
=\frac{1}{2\pi\alpha'}g_{\mu\nu}\dot X^\nu+ \B_{\mu\nu}X'^\nu .
\end{equation}
When ensuring the overall variation of the action is zero, the boundary terms evaluated at the string endpoints ($\sigma = 0, \pi$) must vanish. For freely moving endpoints, this forces the following constraint\footnote{$\mu,\nu$ denote target space indices; in the present open string setup and throughout the paper, when we talk about open strings, they are effectively restricted to the D$p$-brane worldvolume directions, i.e.\ $\mu,\nu = 0,1,\ldots,p$.}
\begin{equation}
g_{\mu\nu}X'^\nu+2\pi\a' \B_{\mu\nu}\dot X^\nu=0,
\qquad \sp=0,\pi .
\end{equation}
This boundary condition intimately links the temporal and spatial variations of the string at its endpoints. We can algebraically invert this relationship to express the temporal velocity $\dot {X} ^\mu$ strictly in terms of the spatial gradient $X'^\nu$
\begin{equation}
\dot X^\mu=-\frac{1}{2\pi\alpha'}(\B^{-1}g)^{\mu}{}_{\nu}X'^\nu .
\end{equation}
Conversely, we can isolate the spatial derivative $X'^\mu$, describing the string's local orientation at the boundary as a function of its velocity
\begin{equation}
X'^\mu=-2\pi\alpha'(g^{-1}\B)^{\mu}{}_{\nu}{\dot{X^\nu}} .
\end{equation}
By substituting this spatial derivative expression back into our earlier formula for the canonical momentum \eqref{eq:tensilemomentum}, we establish a new relationship defining an \textit{effective momentum} driven solely by the velocity. This naturally leads to the identification of the \emph{effective open-string metric}, $G_{\mu \nu}$, which governs the physical dynamics perceived by the open-string in this background
\begin{equation}\label{eq:effectivestringmetric}
\Pi_\mu
= \frac{1}{2\pi\alpha'}\left[g_{\mu\nu}-(2\pi\alpha')^2(\B g^{-1}\B)_{\m\n}\right]\dot X^\n
= \frac{1}{2\pi\a'} G_{\m \n}\dot X^\n ,
\end{equation}
where the explicit matrix form of this effective metric $G_{\mu\nu}$ is given by
\begin{equation}
    G_{\m\n}=\left[g_{\m\n}-(2\pi\a')^2(\B g^{-1} \B)_{\mu\nu}\right],
\end{equation}
and its inverse, $G^{\mu\nu}$, is expressed as
\begin{equation}
    G^{\mu \nu}=\left(\frac{1}{g+2\pi\a' \B}\right)^{\mu\nu}_S = \left(\frac{1}{g+2\pi\alpha'\B}\; g \;\frac{1}{g-2\pi\alpha' \B}\right)^{\mu\nu},
\end{equation}
where $(~)_S$ denotes the symmetric part of the matrix.
\vspace{.2cm}
\paragraph{Symplectic structure.}
Returning to the phase space formulation, the symplectic potential $\theta^b$ (now tracking factors of $\alpha'$) takes the form
\be
 \theta^b = \left(\frac{1}{2 \pi \alpha^\prime} \eta^{ab} g_{\m\n} + \veps^{ab}\B_{\m\n}\right) \pa_a X^\m \d X^\n.
\ee
Taking two antisymmetrised exterior variations provides the symplectic current given by
\begin{align*}
\omega^b(X; \delta_1 X, \delta_2 X) &= \delta_1 \theta^b(X, \delta_2 X) - \delta_2 \theta^b(X, \delta_1 X) \\
  &= \left( \frac{1}{2 \pi \alpha^\prime}\eta^{ab} g_{\m\n} + \veps^{ab}\B_{\m\n} \right)\big( \pa_a \d_1 X^\m \d_2 X^\n - \pa_a \d_2 X^\m \d_1 X^\n \big)\\
  &=\left( \frac{1}{2 \pi \alpha^\prime}\eta^{ab} g_{\m\n} + \veps^{ab} \B_{\m\n} \right) \pa_a \d X^\m \cw \d X^\n .
\end{align*}
To find the global pre-symplectic form $\Omega$, we integrate the temporal component of this current ($\omega^\tau$) across a constant worldsheet time slice (from $\sigma=0$ to $\sigma=\pi$). Notice how the canonical momentum $\Pi_\mu$ naturally factors into the final simplified step
\begin{align*}
\bos{\Omega}&=\int_0^\pi d\sigma\,\omega^\tau \\
&=\int_0^\pi d\sigma\,\left( \frac{1}{2 \pi \alpha^\prime}\eta^{\t\t}g_{\m\n} \, \d X^\m \cw \pa_\tau\d X^\n + \veps^{\sp\t} \B_{\m\n}  \, \d X^\m \cw \pa_\sigma\d X^\n \right)\\
&=\int_0^\pi d\sigma\,\delta X^\mu\cw\delta \Pi_\mu .
\end{align*}
We can parse this integral by isolating the distinct contributions stemming from the symmetric metric $g$ and the antisymmetric $B$-field into $\bos{\Omega}_g$ and $\bos{\Omega}_B$, respectively
\begin{equation}
\bos{\Omega}=\int d\sigma
\left(
\frac{1}{2\pi\alpha'}\,g_{\mu\nu}\,\delta X^\mu\cw\delta\dot X^\nu
+
\B_{\m\n}\,\delta X^\mu\cw\delta X'^\nu
\right)
\equiv \bos{\Omega}_g+\bos{\Omega}_B .
\end{equation}
The $B$-field piece requires a bit of manipulation via integration by parts. Using the product rule for the spatial derivative of the wedge product
\begin{equation}
\partial_\sigma(\delta X^\mu\cw\delta X^\nu)
=
\delta X'^\mu\cw\delta X^\nu
+
\delta X^\mu\cw\delta X'^\nu ,
\end{equation}
and combining this with the innate antisymmetry of $B_{\mu\nu}$, the bulk integral collapses entirely. This localises the $\bos{\Omega}_B$ contribution strictly to the boundary endpoints of the string
\begin{equation}
\bos{\Omega}_B
=
\frac12
\Big[ \B_{\m\n}\delta X^\mu\cw\delta X^\nu \Big]_{0}^{\pi}.
\end{equation}
Next, we examine the metric contribution, $\bos{\Omega}_g$. At the boundaries, we can substitute our previously derived relation linking $\dot {X} ^\mu$ to $X'^\nu$
\begin{equation}
\bos{\Omega}_g
=
-\frac{1}{(2\pi\alpha')^2}
\int g_{\mu\nu}(\B^{-1}g)^{\nu}{}_{\rho} \,
\delta X^\mu\cw\delta X'^\rho .
\end{equation}
Similarly to the $B$-field term, the structural antisymmetry of the composite matrix $(g \B^{-1}g)_{\mu\nu}$ ensures that this integral also evaluates to a pure boundary term
\begin{equation}
\bos{\Omega}_g
=
-\frac{1}{2(2\pi\alpha')^2}
\left[
(g \B^{-1}g)_{\mu\nu} \,
\delta X^\mu\cw\delta X^\nu
\right]_{\partial}.
\end{equation}
\vspace{.05cm}
\paragraph{Boundary symplectic form.}
By superimposing these two evaluated terms, we construct the total symplectic form. This equation explicitly highlights how the background fields couple together to warp the phase space exactly at the string endpoints
\begin{equation}
\bos{\Omega}
= \frac12 \left[ \B_{\m\n} - \frac{1}{(2\pi\a')^2}(g \B^{-1}g)_{\m\n}
\right] \d X^\m\cw\d X^\n .
\end{equation}
The inverse of the symplectic matrix components determines the Poisson bracket of the coordinates. Writing $\bos{\Omega}$ generically as
\begin{equation}
\bos{\Omega}
=
\frac12
\Omega_{\mu\nu}
\delta X^\mu\cw\delta X^\nu .
\end{equation}
The Poisson bracket is given by
\begin{equation}
\{X^\mu,X^\nu\}=\Omega^{\mu\nu}.
\end{equation}
To elegantly compute this inverse, it is mathematically convenient to define an auxiliary matrix $\E$ that bundles the metric and $B$-field together
\begin{equation}
\E=g+2\pi\alpha'\B ,
\end{equation}
Through straightforward matrix algebra, the operator in the total symplectic form can be factored using this newly defined matrix $\E$
\begin{equation}
\B - \frac{1}{(2\pi\alpha')^2}\,g \B^{-1}g
= -(2\pi\a')^{-2} (g+2\pi\a'\B) \B^{-1}(g-2\pi\a' \B).
\end{equation}
Substituting this factored form back into our expression for $\Omega$ gives 
\begin{equation}
\bos{\Omega}
=
-(2\pi\a')^{-2}\E\,\B^{-1}\,\E^T .
\end{equation}
Finally, calculating the inverse of this symplectic matrix provides the fundamental noncommutative parameter $\Theta^{\mu\nu}$
\begin{equation}
\Theta^{\mu\nu}
= 2\pi\a'\left(\frac{1}{g+2\pi\a' \B}\right)^{\m\n}_A.
\end{equation}
where, $(~)_A$ denotes the antisymmetric part of the matrix $G^{\m\n}$.
In matrix form,
\begin{equation}
\mathbf{\Theta}
= -(2\pi\alpha')^2 (g+2\pi\alpha' \B)^{-1} \B (g-2\pi\a' \B)^{-1}.
\end{equation}
Therefore, the classical Poisson brackets (and upon quantisation, the commutators) for the coordinates restricted to the open-string endpoints do not vanish. Instead, they satisfy a non-trivial commutation relation governed by $\Theta^{\mu\nu}$
\begin{equation}
\{X^\mu,X^\nu\}=\Theta^{\mu\nu},
\end{equation}
This profound result is exactly the famous Seiberg-Witten noncommutativity parameter.

\subsection{Tensionless reduction of the noncommutativity parameter}
We now examine how the Seiberg--Witten noncommutativity parameter behaves in the tensionless regime. Physically, this limit corresponds to \(\alpha'\to\infty\), or equivalently \(T\to 0\), one expects the antisymmetric background to dominate over the metric contribution. It is therefore natural to study the large-\(\alpha'\) limit of \(\Theta^{\mu\nu}\) and check whether it reproduces the intrinsic noncommutative structure of the tensionless theory.

The \emph{tensionless} (or Carrollian) string corresponds to 
\begin{equation}
T=\frac{1}{2\pi\alpha'}\longrightarrow 0,
\qquad\Longleftrightarrow\qquad
\alpha'\longrightarrow\infty .
\end{equation}
We now take this limit in the tensile string noncommutativity parameter
\begin{equation}
\mathbf{\Theta}
=-(2\pi\alpha')^2\,
(g+2\pi\alpha'\B)^{-1}\,
\B\,
(g-2\pi\alpha'\B)^{-1}.
\end{equation}
For notational convenience, define
\begin{equation}
\chi \equiv 2\pi\alpha' .
\end{equation}
Then
\begin{equation}
\mathbf{\Theta}
=
-\chi^2\,
(g+\chi \B)^{-1}\,
\B\,
(g-\chi\B)^{-1}.
\end{equation}

To extract the large-\(\chi\) behaviour, factor out \(\chi\) from each matrix appearing inside the inverses. First,
\begin{equation}
g+\chi\B= \chi\left(\B+\frac{1}{\chi}g\right)  \implies 
(g+\chi\B)^{-1}=\frac{1}{\chi} \left(\B+\frac{1}{\chi}g\right)^{-1}.
\nonumber
\end{equation}
Similarly,
\begin{equation}
g-\chi\B =-\chi\left(\B-\frac{1}{\chi}g\right) \implies (g-\chi\B)^{-1}
= -\frac{1}{\chi} \left(\B-\frac{1}{\chi}g\right)^{-1}. \nn
\end{equation}
Substituting these two expressions into \(\mathbf{\Theta}\), we obtain
\begin{equation}
\begin{split}
\mathbf{\Theta}
&=
-\chi^2
\left[
\frac{1}{\chi}
\left(\B+\frac{1}{\chi}g\right)^{-1}
\right]
\B
\left[
-\frac{1}{\chi}
\left(\B-\frac{1}{\chi}g\right)^{-1}
\right]
\\[4pt]
&=
\left(\B+\frac{1}{\chi}g\right)^{-1}
\B
\left(\B-\frac{1}{\chi}g\right)^{-1}.
\end{split}
\end{equation}
Now take the limit \(\chi\to\infty\). Since \(\chi^{-1}\to 0\), we have
\begin{equation}
\left(\B+\frac{1}{\chi}g\right)^{-1}
\longrightarrow
\B^{-1},
\qquad
\left(\B-\frac{1}{\chi}g\right)^{-1}
\longrightarrow
\B^{-1},
\end{equation}
provided \(\B\) is invertible on the relevant brane directions. Therefore,
\begin{equation}
\mathbf{\Theta}
\longrightarrow
\B^{-1}\,\B\,\B^{-1}
=
\B^{-1}.
\end{equation}
Thus the large-\(\alpha'\) limit of the Seiberg--Witten noncommutativity parameter is
\begin{equation}
\mathbf{\Theta}_{\rm tensionless}
=
\lim_{\alpha'\to\infty}\mathbf{\Theta}
=
\B^{-1}
.
\end{equation}

Equivalently, one may write the same result as an asymptotic expansion. Using
\begin{equation}
\left(\B\pm \frac{1}{\chi}g\right)^{-1}
=
\B^{-1}
\mp
\frac{1}{\chi}\B^{-1}g\B^{-1}
+
O(\chi^{-2}),
\end{equation}
one finds
\begin{equation}
\begin{split}
\mathbf{\Theta}
&=
\left[
\B^{-1}
-
\frac{1}{\chi}\B^{-1}g\B^{-1}
+
O(\chi^{-2})
\right]
\B
\left[
\B^{-1}
+
\frac{1}{\chi}\B^{-1}g\B^{-1}
+
O(\chi^{-2})
\right]
\\[4pt]
&=
\B^{-1}
+
O(\chi^{-1})
=
\B^{-1}
+
O\!\left(\frac{1}{\alpha'}\right).
\end{split}
\end{equation}
Hence, the noncommutativity parameter has a smooth and finite tensionless limit.

In the next section, we derive this same noncommutativity parameter intrinsically from the CPS of the tensionless open string itself. The resulting boundary Poisson structure agrees precisely with the tensionless reduction above, thereby showing that the intrinsic tensionless noncommutativity parameter coincides with the \(\alpha'\to\infty\) limit of the usual tensile-string expression.

\section{CPS for tensionless strings}\label{sec:cps_tensionless}
 Now, we will apply the CPS formulation \cite{Crnkovic:1986ex, Iyer:1994ys, Harlow:2019yfa} to the tensionless string theory. We will mainly follow the IW procedure discussed in section \ref{sec:cps}. We consider the intrinsically tensionless string described by the Isberg-Lindstrom-Sundborg-Theodoridis (ILST) action \cite{Isberg:1993av}
\begin{equation}\label{eq:ILST}
S_0[X,V]
= \int_{\Sigma} d^2\sp\;
V^a V^b\, \pa_a X^\m \pa_b X_\m ,
\end{equation}
where $X^\mu(\sigma)$ are target-space embedding coordinates and $V^a$ is a worldsheet vector density. The abstract Roman indices $a, b,\dots$ run in the worldsheet coordinates $\t$ and $\sp$; whereas the Greek indices $\m, \n, \dots$ run in the spacetime coordinates. 

For open strings, the worldsheet $\Sigma$ has a boundary $\partial\Sigma$, and the variational principle is defined by requiring that the total boundary variation of the action vanish under suitable boundary conditions. The variation of the action \eqref{eq:ILST} yields
\begin{align}\label{variation_ILST}
    \delta S_0 = \, 2&\int_\Sigma d^2\sigma\; \left[ -\pa_a\left(V^a V^b \pa_b X_\m \right)\delta X^\m  +  \left(V^a \pa_a X^\m \pa_b X^\n \eta_{\m\n}\right)\delta V^b\right] \nonumber \\
    + 2 &\int_{\Sigma} d^2\sp \, \pa_a \left[ \left(V^a V^b \pa_b X_\m\right) \delta X^\m\right]\,.
\end{align}
The second integral is a total derivative and hence the boundary term. We can rewrite the above variation in terms of differential forms as follows
\begin{equation}\label{variation_lagrangian_ILST}
\d \bos{L}_0
= \bos{E}_\mu\,\delta X^\mu + \bos{E}_b\,\delta V^b + \ed \bost_0(X^\m,\delta X^\m),
\end{equation}
where $\bos{L}_0$ denotes the Lagrangian $2$-form in the worldsheet, $\bos{E}$ denotes the equations of motion
\begin{align} 
    \bos{E}_\m &= 0 \implies \pa_a\left(V^a V^b \pa_b X_\m \right)=0 \\
    \bos{E}_b &= 0 \implies V^a \g_{ab}=0
\end{align}
where we have used the definition of the induced metric of the worldsheet: $\g_{ab}=\pa_a X^\m \pa_b X^\n \eta_{\m\n}$ and $\bost_0$ is the symplectic potential $1$-form. The suffix `$0$' implies that the theory is free in the sense that it does not have any other field. Later, we will also look at tensionless strings in the presence of a Kalb-Ramond field and its CPS formulation. From \eqref{variation_ILST} and \eqref{variation_lagrangian_ILST}, we can easily identify the symplectic potential given by
\begin{equation}\label{symplectic_potential}
\theta_0^a (X, \delta X)
=2V^a V^b\,\partial_b X_\mu\,\delta X^\mu .
\end{equation} 
We can define a quantity like 
\be
P^a_\m = \dfrac{\d L_0}{\d (\pa_a X^\m)}= 2 V^a V^b \pa_b X_\m,
\ee
which is an analogue of the $d$-dimensional momentum density on the worldsheet \cite{Green:1987sp} in the tensionless limit, and $\bos{L}_0=*L_0$ with `$*$' being the Hodge dual. Then, the symplectic potential can be written as
\be
\theta_0^a = P^a_\m \d X^\m.
\ee
Now, we can obtain the Noether current form as defined in \eqref{noethercurrent}, which yields
\begin{align*}
     J^a_{0_\xi} &= \theta^a_0 (X,\L_\xi X)- \xi^a \bos{L}_0 \\
     &= 2V^a V^b\,\partial_b X_\m\,\L_\xi X^\m - \xi^a V^b V^c\, \pa_b X_\m \pa_c X^\m \\
     &= -2V^a V^b\,\partial_b X_\m\,\xi^c \pa_c X^\m- \xi^a V^b V^c\, \pa_b X_\m \pa_c X^\m \\
     &=-(2V^a\xi^c+\xi^a V^c)\, \underbrace{V^b\,\partial_b X_\m \pa_c X^\m}_{V^b\, \gamma_{bc}} .
\end{align*}
In the third line, we have used the transformations $\L_\xi  X^\m = - \xi^a \pa_a X^\m $ \cite{Isberg:1993av, Bagchi:2026wcu}, with $\L_\xi$ being the Lie derivative w.r.t the worldsheet diffeomorphism generator $\xi^a$. Using the e.o.m., $V^b\, \gamma_{bc}=0$, we find that the Noether current vanishes identically on-shell,
\begin{equation}
J^{a}_{0_\xi}\;\approx\;0.
\end{equation}
Here `$\approx$' denotes that the quantity is derived on-shell. The vanishing of the Noether current implies that the intrinsically tensionless string admits no nontrivial
\emph{bulk} Noether charge. Any conserved quantity must therefore arise
from boundary contributions to the action. This observation will become
crucial once a Kalb-Ramond background is included, where a nontrivial
boundary symplectic structure emerges.

To obtain the symplectic form for the tensionless closed string theory, we first need to compute the symplectic current as defined in \eqref{sympcurrent}. As the action \eqref{eq:ILST} is worldsheet diffeomorphism invariant, we can also use the formula \eqref{sympcurrent_diffs} to find the symplectic current. 

 The standard way to study tensionless strings (and tensile strings) is to fix a gauge and then take a look at it. So, we will compute the symplectic potential after fixing the vector densities $V^a$.
In the ILST action, the vector densities are defined via the Lagrange multipliers of the constrained Hamiltonian \cite{Isberg:1993av},
\be\label{eq:vector_density_definition} 
  V^a = \dfrac{1}{\sqrt{\l}}\, \left(1, -\rho\right)\,.
\ee
Now, one can use different gauges \cite{Isberg:1993av, Bagchi:2026wcu, Bagchi:2015nca} depending on which limit of the tensionless string one intends to study. But, in every case, the vector densities become constant under any perturbation. So, we can proceed with the following assumption
\be
 \d V^a = 0\,.
\ee
This assumption simplifies the calculations as the only remaining dynamical fields are the $X^\m$'s. We use \eqref{sympcurrent} to compute the symplectic current 
\begin{align*}
    \omega_0^a (X, \d_1 X, \d_2 X) &= \d_1 \theta_0^a (X, \d_2 X) - \d_2 \theta_0^a (X, \d_1 X) \\
    &= 2 V^a V^b \Big( \d_2 X^\m \,\pa_b \,\d_1 X^\n - \d_1 X^\m\,\pa_b\,\d_2 X^\n \Big) \eta_{\m\n}\\
    &= 2 V^a V^b\eta_{\m\n}\, \pa_b \, \d_1 X^\m \cw \d_2 X^\n \\
    &= \d_1 P_\n^a \cw \d_2 X^\n.
\end{align*}
Here, we have used `$\cw$' to denote the wedge product in the field space. For the worldsheet or spacetime wedge product, we will use the symbol `$\w$',\footnote{In the above derivation and in the rest of the article, we are using the following notation for wedge product \cite{Wald:1984rg},
\be
 \bos{\a} \w \bos{\b} = \bos{\a} \otimes \bos{\b} - \bos{\b} \otimes \bos{\a}\,, \nn
\ee
for any two forms $\bos{\a}$ and $\bos{\b}$, and `$\otimes$' being the tensor product. The same goes for the field space wedge product. 
\label{wedge_product}} as mentioned in section \ref{sec:cps}.
We fix the gauge to $V^a = (v,0)$, the momentum at the boundary is found to vanish, $P^a_\mu = 0$ (as a boundary condition) \cite{Duary:2025hdb}. Consequently, the momentum is zero, which in turn implies that the symplectic form vanishes,
\be
 \bom_0  (X, \d_1 X, \d_2 X) = 0.
\ee
 The symplectic form is defined by integrating the symplectic current over a Cauchy surface. The result obtained above shows that this symplectic form also vanishes
\be
 \bos{\Omega}_0  (X, \d_1 X, \d_2 X) = \int_\C \bom_0  (X, \d_1 X, \d_2 X) =0.
\ee
Therefore, the symplectic form is degenerate and hence is a pre-symplectic form. As the gauge fixed (or reduced) CPS does not carry a nontrivial symplectic structure, it implies that the tensionless strings do not admit any intrinsic Poisson structure.

In this case, the pre-symplectic form vanishes for any general variation of the fields. So, all the bulk variations correspond to the zero modes of $\bos{\Omega}_0$. The generators of these zero modes form a Lie algebra \cite{Harlow:2019yfa}.

\subsection{Presence of a uniform Kalb--Ramond field} \label{sec:cps_tensionless_with_B}
We begin with the Polyakov-type action for a bosonic string coupled to a Kalb--Ramond field,
\begin{equation}
\begin{split}
S_{\text{tensile}} 
&= - \frac{T}{2} \int_\Sigma d^2\sp\left( 
\sqrt{-g}\, g^{ab} \,\partial_a X^\mu \partial_b X^\nu \,\eta_{\mu\nu} 
+ \varepsilon^{ab} B_{\mu\nu} \,\partial_a X^\mu \partial_b X^\nu 
\right).
\end{split}
\label{eq:polyakov_action}
\end{equation}
A useful way to formulate the tensionless limit is to express it in a dimensionless manner. Since \(\alpha'\) is dimensionful, one should not simply say \(\alpha'\to\infty\); rather, one introduces a fixed reference scale \(c'\) of dimension \((\text{length})^2\) and writes
\begin{equation}
\alpha'=\frac{c'}{\epsilon}, \qquad \epsilon\to 0.
\end{equation}
Then
\begin{equation}
T=\frac{1}{2\pi\alpha'}=\frac{\epsilon}{2\pi c'}\to 0,
\end{equation}
so the tensionless limit is governed by the dimensionless parameter \(\epsilon\). Equivalently, the intrinsic string length grows as \(\ell_s\sim \sqrt{\alpha'}\sim \sqrt{c'}\epsilon^{-1/2}\), showing that the limit corresponds to a long-string, vanishing-tension regime. Written this way, the scaling makes transparent that one is really taking the ratio \(\alpha'/c'\) to be large, while keeping \(c'\) fixed. A nontrivial limit then requires the background couplings to be rescaled simultaneously (see eq.\eqref{comres}) so that the action remains finite.

To obtain a well-defined tensionless theory, we consider the limit in which the string tension vanishes. To retain nontrivial dynamics, the worldsheet fields must be redefined simultaneously. Specifically, we perform the following replacements
\begin{equation}
\label{rescal1}
\begin{split}
-\dfrac{T}{2} \sqrt{-g}\, g^{ab} &\longrightarrow V^a V^b, \\
B_{\mu\nu} &\longrightarrow - \frac{B_{\mu\nu}}{\epsilon},
\end{split}
\qquad\text{with}\qquad
V^a = \frac{1}{\sqrt{\lambda}} (1, -\rho).
\end{equation}
Here, $V^a$ is interpreted as a timelike vector density (or vielbein-like object) on the worldsheet, encoding the degenerate Carrollian geometry characteristic of tensionless strings as done in the free background case. Notice that we have an extra minus sign in the $B$-field as opposed to \cite{Banerjee:2024fbi, Duary:2025hdb}, because our Polyakov action differs by a minus from those.
To keep the antisymmetric background field finite in this limit, we introduce a rescaled field $\B_{\mu\nu}$ via\footnote{Notice that our rescaling is different from the existing literature \cite{Banerjee:2024fbi, Duary:2025hdb}.}
\begin{equation}
\begin{split}
\B_{\mu\nu} &= \frac{1}{2\pi \alpha'} B_{\mu\nu}. \\
\end{split}
\label{eq:rescaled_B}
\end{equation}
Under this redefinition, the $B$-field contribution remains finite as $T \to 0$.
The physical reason for the scaling of the Kalb--Ramond field is that, in the tensile action, the antisymmetric background does not appear as an independent coupling, but only through the combination multiplied by the string tension. Hence, if one takes the tensionless limit \(T\to 0\) while keeping \(B_{\mu\nu}\) fixed, the entire antisymmetric coupling vanishes, and the background decouples from the worldsheet dynamics. 
Here, 
\begin{equation}
\label{comres}
\B_{\mu\nu} =-\frac{1}{2\pi c'} B_{\mu\nu}. 
\end{equation}
remains finite as \(T\to 0\) or equivalently \(\alpha'\to\infty\).

Substituting these scalings into the original action and taking the strict tensionless limit yields the action for an intrinsically tensionless string in a constant Kalb--Ramond background
\begin{equation}
\begin{split}
S_{\text{tensionless}}&= \int_\Sigma d^2\sp \left( 
V^a V^b \,\partial_a X^\mu \partial_b X^\nu \,\eta_{\mu\nu} 
+ \dfrac{1}{2} \varepsilon^{ab} \,\B_{\mu\nu} \,\partial_a X^\mu \partial_b X^\nu 
\right).
\end{split}
\label{eq:tensionless_action_B}
\end{equation}
This action generalises the original ILST formulation, which describes free tensionless strings in flat spacetime by consistently incorporating coupling to a constant antisymmetric tensor field. It will serve as the starting point for the CPS analysis that follows.

Now, we will compute the symplectic form for the tensionless string in a constant Kalb--Ramond background, using the CPS formalism.
From the total action \eqref{eq:tensionless_action_B}, the variation yields
\begin{equation}
\d S_{\text{tensionless}} = \d S_0 + \d S_B\,, \nn
\end{equation}
where $\d S_0$ is given in \eqref{variation_ILST} and $\d S_B$ is given by,
\be
 \d S_B = \int_\Sigma d^2 \sp \left[ -\varepsilon^{ab}  \B_{\m\n} \d X^\m \pa_a\pa_b X^\n +  \pa_a \left(  \varepsilon^{ab} \B_{\m\n} \d X^\m \pa_b X^\n \right) \right].
\ee
Combining it with the ILST action and rewriting in the form language, we get
\be
 \d \bos{L} = \bos{E}_\m\,\d X^\m + \bos{E}_b\,\d V^b + \ed \bost(X^\m,\d X^\m),
\ee
where the e.o.m.s are given by
\begin{subequations}
\be
  \bos{E}_\m = 0 \implies \pa_a\left(V^a V^b \eta_{\m\n} \pa_b X^\m \right) =0 ; 
\ee
 and for the vector density $V^a$,
 \be
 \bos{E}_b = 0 \implies V^a \g_{ab}=0 \nn .
 \ee
\end{subequations}
The symplectic potential is given by 
\be
\theta^a (X, \delta X) = \theta^a_0 +\theta^a_B =\left(2\eta_{\m\n} V^a V^b + \varepsilon^{ab} \B_{\m\n}\right) \,\d X^\m\,\pa_b X^\n .
\ee 
The (pre)symplectic current is defined by
\begin{equation}
\omega^a(X; \delta_1 X, \delta_2 X) = \delta_1 \theta^a(X, \delta_2 X) - \delta_2 \theta^a(X, \delta_1 X). \nn
\end{equation}
However, the pre-symplectic current vanishes identically for the ILST action, as we saw in the previous section. Here, we will compute the same for the Kalb-Ramond part of the action as follows
\begin{equation}\label{eq:symp_current_B}
\begin{split}
 \omega^a_B 
&= \, \varepsilon^{ab} \B_{\mu\nu} 
   \big( \d_1 X^\m \,\pa_b\d_2 X^\n -  \d_2 X^\m \,\pa_b\d_1 X^\n \big) \nn\\
&=  \dfrac{1}{2}\varepsilon^{ab} \B_{\mu\nu} \pa_b \big( \d_1 X^\m \cw \d_2 X^\n \big) \nn\\
&= \pa_b \left(\dfrac{1}{2}\varepsilon^{ab} \B_{\mu\nu} \,\d_1 X^\m \cw \d_2 X^\n \right),
\end{split}
\end{equation}
where we used the fact that $\B_{\mu\nu}$ and $\varepsilon^{ab}$ are constant. Thus, the $\B$-contribution to the (pre)symplectic current is exact
\begin{equation}
\omega^b_B = \pa_b k^{ab}_{B}, \qquad 
k_B^{ab} = \dfrac{1}{2} \varepsilon^{ab} \B_{\mu\nu} \,\delta X^\m \cw \delta X^\n .
\end{equation}
The total (pre)symplectic current is
\begin{equation}
\omega^a = \omega^a_0 + \omega^a_B = \partial_b k^{ab}.
\end{equation}
The (pre)symplectic form on a Cauchy slice $\C$ (e.g., at fixed $\tau$) is obtained by integrating $\omega^\tau$ over $\sigma$:
\begin{equation}
\begin{split}
\bos{\Omega}
&= \int_\C d\sigma\, \omega^\tau 
 = \int_\C d\sigma\, \partial_\sigma k^{\tau\sigma} \\
&= k^{\tau\sigma}\big|_{\sigma=\pi} - k^{\tau\sigma}\big|_{\sigma=0} 
 = \Big[ k^{\tau\sigma}\Big]_{\partial \mathcal{C}}.
\end{split}
\end{equation}
Using $\varepsilon^{\tau\sigma} = +1$, we have
\begin{equation}
k^{\tau\sigma} =\dfrac{1}{2} \B_{\mu\nu} \,\d X^\m \cw \d X^\n 
               =\dfrac{1}{2} \B_{\n\m} \,\d X^\n \cw \d X^\m .
\end{equation}
Therefore, the total (pre)symplectic form is purely boundary-supported
\begin{equation}
\bos{\Omega} =\left[\dfrac{1}{2} \B_{\mu\nu} \,\delta X^\mu \cw \delta X^\nu  \right]_{\partial \mathcal{C}}.
\end{equation}
For an open-string with endpoints at $\sigma = 0$ and $\sigma = \pi$, this becomes
\begin{equation}
\bos{\Omega} =\dfrac{1}{2} \B_{\mu\nu} \Big[ \d X^\m(\tau,\pi) \cw \d X^\n(\tau,\pi) 
 - \d X^\m(\tau,0) \cw \d X^\n(\tau,0)
\Big].
\end{equation} 
The reduced boundary symplectic form is
\begin{equation}\label{symp_form_tensionless_with_B}
\bos{\Omega}_{\mathrm{bdry}} = \dfrac{1}{2}\B_{\mu\nu} \,\d X^\m \cw \d X^\nu .
\end{equation}

\subsubsection*{Emergence of noncommutativity}\label{sec:noncomm_tensionless}
As we have a symplectic form in hand, it can be used to define the \emph{Poisson bracket} as
\be
 \{ f, h \} \equiv \bos{\Omega}^{-1}(\d f, \d h) = \Omega^{\m\n}\, \hdel_\m f \hdel_\n h \,,
\ee
where $\{,\}$ denotes the Poisson bracket of two smooth functions $f$ and $h$ on the symplectic manifold, `$\hdel$' denotes the partial derivative on it, and $\Omega^{\m\n}$ is the inverse of the symplectic form (for more details see Appendix \ref{sec:symp_mech}). Applying this definition to $X^\m$'s yields the following Poisson bracket
\be
\{X^\m , X^\n \} = \Omega^{\m\n}.
\ee
 The r.h.s. is non-zero given a non-vanishing symplectic form. In that case, the background geometry becomes noncommutative and the r.h.s. is related to the noncommutative parameter $\Theta^{\m\n}$. For the tensionless string with a uniform background Kalb-Ramond field, the symplectic form is given by  \eqref{symp_form_tensionless_with_B}. So, the Poisson bracket of the target space becomes
\be
 \{X^\m , X^\n \} = \B^{\m\n}.
\ee
The boundary symplectic structure makes the geometry noncommutative. It implies that the tensionless strings in the presence of a uniform Kalb-Ramond field have an intrinsically noncommutative geometry with the noncommutative parameter
\be
 \Theta^{\m\n} = \B^{\m\n}.
\ee
This derivation of the noncommutative structure of the boundary phase space of tensionless strings is inherently covariant and bypasses the first-order Lagrangian method of \cite{Duary:2025hdb}.

\subsection{Including boundary gauge field}\label{sec:cpsb_bdry_gauge-field}

In this section, we incorporate a boundary gauge field term using the CPSB framework. We start with the Polyakov action for a bosonic string ending on a D$p$-brane, including the boundary gauge field term
\begin{equation}
    S^A_{\text{tensile}} =- \frac{1}{4\pi\alpha'} \int_\Sigma d^2\sigma \left[
\sqrt{-g} g^{ab} g_{\m\n} \pa_a X^\m \pa_b X^\n 
+ \veps^{ab} B_{\m\n} \pa_a X^\m \pa_b X^\n 
\right] 
+ \frac{1}{2\pi\alpha'} \oint_{\partial\Sigma} d\tau \, A_i(X) \partial_\tau X^i
\label{eq:full-action-bdry}
\end{equation}
where $A_i$ ($i=0, 1, \cdots, p$) is the $U(1)$ gauge field living on the D$p$-brane. We use the convention $\eta^{ab} = \text{diag}(-1, 1)$ and $\veps^{\t\sp} = 1$. The string background is taken to be flat with $g_{\mu\nu} = \eta_{\mu\nu}$, and $\bos{H} = \ed\bos{B} = 0$.
Using the string tension $T = \frac{1}{2\pi\alpha'}$, the boundary term can be written as
\begin{equation}
S_{\text{bdry}} = T \oint_{\partial\Sigma} d\tau \, A_i(X) \dot{X}^i
\label{eq:boundary-action-T}
\end{equation}
where $\dot{X}^i \equiv \partial_\tau X^i$.
To obtain the tensionless theory, we apply the scaling limits. Specifically, we take $T \to 0$ while rescaling the fields to keep the dynamics finite
\begin{align}
-\frac{T}{2}\sqrt{-g} g^{ab} &\longrightarrow V^a V^b, \nn\\
B_{\mu\nu} &\longrightarrow -\frac{\B_{\mu\nu}}{\epsilon}, \quad \epsilon \to 0,\nn \\
A_i &\longrightarrow \frac{\mathcal{A}_i}{\epsilon}, \quad \text{such that } T A_i \to \dfrac{1}{2} \mathcal{A}_i.
\label{eq:tensionless-scaling}
\end{align}
Here, $V^a$ is the worldsheet vector density
\begin{equation}
V^a = \frac{1}{\sqrt{\lambda}}(1, -\rho) \nn
\label{eq:vector-density}
\end{equation}
and $\mathcal{A}_i$ is the rescaled finite gauge field. The tensionless action becomes
\begin{equation}
S_{\text{tensionless}}^{\mathcal{A}}= \int_\Sigma d^2\sigma \left( V^a V^b \partial_a X^\mu \partial_b X^\nu \eta_{\mu\nu} + \dfrac{1}{2} \veps^{ab} \B_{\mu\nu} \partial_a X^\mu \partial_b X^\nu \right) 
+ \dfrac{1}{2} \oint_{\partial\Sigma} d\tau \, \mathcal{A}_i(X) \dot{X}^i
\label{eq:tensionless-action-bdry}
\end{equation}
Following the CPSB formalism, we identify the boundary Lagrangian form $\bos{\ell}$ as
\begin{equation}
\bos{\ell} = \dfrac{1}{2} \mathcal{A}_i(X) \dot{X}^i \, d\tau.
\label{eq:boundary-lagrangian}
\end{equation}
The variation of the action includes boundary terms from both the bulk symplectic potential $\bost_{\text{bulk}}$ and the variation of the boundary Lagrangian $\delta \ell$. We get
\begin{equation}
(\bost_{\text{bulk}} + \delta \bos{\ell})\big|_{\pa\Sigma} = \ed\bos{\sc}
\label{eq:stationarity}
\end{equation}
for any $(d-2)$-form $\bos{\sc}$. We compute the variation of $\ell$ with respect to the field space exterior derivative $\delta$
\begin{align}
\d \bos{\ell} &= \dfrac{1}{2}\delta \left( \mathcal{A}_i \dot{X}^i \right) d\tau \nonumber \\
&= \dfrac{1}{2}\left( \delta \mathcal{A}_i \, \dot{X}^i + \mathcal{A}_i \, \delta \dot{X}^i \right) d\tau. \nonumber 
\label{eq:delta-l-1}
\end{align}
Using $\delta \mathcal{A}_i = \partial_j \mathcal{A}_i \, \delta X^j$, the variation becomes
\begin{align}
\d \bos{\ell}
&= \dfrac{1}{2}\left( \partial_j \mathcal{A}_i \, \delta X^j \dot{X}^i 
+ \mathcal{A}_i \, \delta \dot{X}^i \right) d\tau .
\end{align}
The second term can be integrated by parts along the boundary worldline,
\begin{align}
\mathcal{A}_i \, \delta \dot{X}^i
= \partial_\tau (\mathcal{A}_i \delta X^i)
- (\partial_j \mathcal{A}_i) \dot{X}^j \delta X^i .\nn 
\end{align}
Substituting this back, we obtain
\begin{align}
\d\bos{\ell}
&= \left[ \dfrac{1}{2}
(\partial_j \mathcal{A}_i - \partial_i \mathcal{A}_j)
\dot{X}^i \delta X^j
\right] d\tau
+ \partial_\tau \left(\dfrac{1}{2} \mathcal{A}_i \delta X^i \right) d\tau .
\end{align}
The antisymmetric combination defines the field strength
\begin{equation}
\ff_{ij} = \pa_i \A_j - \pa_j \A_i .
\end{equation}
Therefore
\begin{equation}
\d \bos{\ell}
= \dfrac{1}{2} \ff_{ij} \dot{X}^i \d X^j d\tau 
+ \ed\left(\dfrac{1}{2} \A_i \d X^i \right) .
\end{equation}
Comparing with the CPSB stationarity condition
\begin{equation}
(\bos{\theta}_{\rm bulk} + \d \bos{\ell})|_{\pa\Sigma} = \ed \bos{\sc} , \nn
\end{equation}
we see that the total derivative term can be absorbed into $\bos{\sc}$,
\begin{equation}
\bos{\sc} =\dfrac{1}{2} \mathcal{A}_i \delta X^i .
\end{equation}
The bulk action follows the same derivation as was done in section \ref{sec:cps_tensionless_with_B}, and the symplectic potential is given by
\be
\theta^b (X, \delta X) =\left(2\eta_{\m\n} V^a V^b + \varepsilon^{ab} \B_{\m\n}\right) \,\partial_a X^\m\,\delta X^\n . \nn 
\ee
As we have both the symplectic potential and the codimension-2 form $\bos{\sc}$, the (pre)-symplectic current is defined as in \eqref{eq:symp_current_HW},
\be
 \omega^b = \d \left(\theta^b - \pa_a \sc^{ab}\right) .
\ee
The first term follows directly from equation \eqref{eq:symp_current_B},
\be
 \d \theta^b = \pa_a \left( \dfrac{1}{2}\veps^{ab} \B_{\m\n} \d X^\m \cw \d X^\n \right).
\ee
Now, the variation of the codimension-2 quantity yields
\begin{align}
 \d \bos{\sc} &= \dfrac{1}{2}\d (\A_i \d X^i) \nn \\
     &= \dfrac{1}{2}\pa_j \A_i \d X^i \cw \d X^j \nn \\
     &= - \dfrac{1}{2}\pa_i \A_j \d X^i \cw \d X^j \nn \\
     &= - \dfrac{1}{2} \ff_{ij} \d X^i \cw \d X^j \nn \\
     &= - \dfrac{1}{2} \ff_{\m\n} \d X^\m \cw \d X^\n .
\end{align}
In the last line, we have changed the indices for later convenience; this does not affect the quantities, since for $\m>p$ (for D$p$-brane) the gauge field itself vanishes. Combining these, we obtain the symplectic current,
\begin{align}
\omega^b &= \d \left(\theta^b - \pa_a \sc^{ab}\right) \nn \\
    &= \pa_a \left( \dfrac{1}{2}\veps^{ab} \B_{\m\n} \d X^\m \cw \d X^\n \right) + \pa_a \left(\dfrac{1}{2} \veps^{ab} \ff_{\m\n} \d X^\m \cw \d X^\n \right) \nn \\
    &= \pa_a k^{ab},
\end{align}
with
\be
 k^{ab} = \dfrac{1}{2} \veps^{ab} \sf_{\m\n} \d X^\m \cw \d X^\n .
 \ee
Here, $\bos{\sf} = \bos{\B}+\bos{\ff}$ is an analogue of the modified Born-Infeld field strength\footnote{The modified Born-Infeld field strength was defined when both ends of the string end on the same D$p$-brane \cite{Chu:1999gi}, as discussed in appendix \ref{sec:alternative_gauge_derivation}. We are using the same notation to denote both.} in the tensionless limit. We find that adding a boundary gauge term also yields the symplectic current as a pure boundary term, similar to the Kalb-Ramond field case. Therefore, the symplectic form is given by
\be
 \bos{\Omega} = \int_{\C} \omega^b. 
\ee
Fixing $\veps^{\tau\sp}=1$, we can write the symplectic form in components as
\be\label{eq:symp_form_tensionless_with_F}
 \bos{\Omega}_{\rm{bdry}} = \left[ \dfrac{1}{2}\sf_{\m\n}\, \d X^\m\cw\d X^\n \right]_{\partial \mathcal{C}} .
\ee
This has the same structure as \eqref{symp_form_tensionless_with_B} with the Kalb-Ramond field replaced by the new field strength $\sf_{\m\n}$. Thus, the gauge field living on the D$p$-brane induces an additional
boundary symplectic structure controlled by its field strength $\ff_{\m\n}$.

\paragraph{Noncommutativity.} Including the boundary gauge term in the action for a bosonic string ending on D$p$-brane yields a boundary symplectic structure in the tensionless limit as depicted in \eqref{eq:symp_form_tensionless_with_F}. We can follow the same procedure as was done in section \ref{sec:noncomm_tensionless}. In the present case, the Poisson bracket between the target space coordinates is given by
\be
 \{ X^\m , X^\n \} = \sf^{\m\n}.
\ee
So, the emergent geometry is noncommutative, similar to the constant $B$-field case. It is easy to identify the noncommutative parameter
\be
 \Theta^{\m\n} = \sf^{\m\n}.
\ee

\section{Conclusions and outlook}
\label{sec:cons}
In this paper, we have developed a CPS analysis of open bosonic strings in constant antisymmetric backgrounds, treating the ordinary tensile theory and its intrinsically tensionless counterpart within a single geometric framework. The main result is that the origin and role of noncommutativity can be understood directly from the symplectic structure, without relying on the standard operator derivation from worldsheet propagators.

For the tensile open-string in a constant Kalb-Ramond background, we showed that the pre-symplectic current decomposes into a nontrivial bulk contribution coming from the kinetic term and an exact boundary contribution generated by the \(B\)-field. Upon imposing the mixed open-string boundary conditions, the full symplectic form reduces at the endpoints to the inverse structure whose antisymmetric part reproduces the Seiberg--Witten noncommutativity parameter. In this sense, the familiar noncommutative geometry on the D-brane worldvolume emerges as a direct consequence of the boundary symplectic data of the CPS.

The tensionless theory exhibits a qualitatively different structure. In the absence of background fields, the bulk pre-symplectic form is degenerate, reflecting the collapse of the ordinary propagating phase space in the Carrollian regime. Once a constant Kalb--Ramond field is turned on, however, the symplectic current becomes exact and localises completely on the boundary. The resulting reduced phase space is therefore supported entirely at the string endpoints, and the corresponding Poisson brackets imply an intrinsic noncommutative geometry,
$$
\{X^\mu, X^\nu\} = \B^{\mu\nu}.
$$
This sharply distinguishes the tensionless regime from the tensile one: in the former, noncommutativity is not merely a boundary deformation of an otherwise nondegenerate bulk phase space, but rather the defining remnant of the physical phase space itself.

We further incorporated a boundary \(U(1)\) gauge-field coupling using the CPS formalism with boundaries. The Harlow--Wu framework makes it possible to treat the boundary term systematically. The resulting pre-symplectic current is again exact, and the effective field-strength combination on the brane controls the boundary symplectic form. Thus, the endpoint noncommutativity in the tensionless theory is governed not only by the Kalb-Ramond field, but more generally by the gauge-invariant antisymmetric data induced on the boundary.\\

Our analysis opens several avenues for future research.

\paragraph{Non-constant antisymmetric backgrounds.}
A natural extension of the present work is to go beyond constant antisymmetric backgrounds and study the covariant phase space in the presence of a non-constant Kalb--Ramond field with non-vanishing field strength \(H_{\mu\nu\rho}=\partial_{[\mu}B_{\nu\rho]}\) \cite{Herbst:2001ai}. See cf. \cite[section 4]{Seiberg:1999vs} for a slowly varying field. In such situations, the pre-symplectic current need no longer remain exact, and genuine bulk contributions to the symplectic structure may survive even in the tensionless limit. A concrete question is whether the resulting reduced phase space continues to be purely boundary-supported or instead acquires a mixed bulk-boundary character.

\paragraph{Deformation quantisation from the boundary phase space.}
An important direction is to quantise the boundary symplectic structure obtained in the present analysis and relate it directly to deformation quantisation \cite{Schomerus:1999ug}. In the standard open string description, the antisymmetric part of the boundary propagator gives rise to the Moyal-Weyl star product and the Seiberg-Witten noncommutativity parameter. By contrast, in the CPS approach, the same structure emerges from the inverse of the boundary symplectic form. It would therefore be very interesting to develop a direct quantisation of the CPS boundary phase space and show how the corresponding noncommutative star product arises purely from the symplectic data, both in the tensile theory and in the intrinsically tensionless regime.

\paragraph{Higher structures and nonassociativity.}
A closely related question is whether the covariant phase space approach can be generalised so as to capture not only ordinary Poisson brackets but also genuinely higher structures. The standard open string story leads to a noncommutative but associative product, whereas more general flux backgrounds in string theory suggest quasi-Poisson, tri-product, and even nonassociative structures \cite{Blumenhagen:2014sba, Cornalba:2001sm, Herbst:2003we}. It is therefore natural to ask whether a suitable extension of CPS or CPSB can detect the geometric origin of such higher brackets, for instance, through a controlled failure of exactness of the pre-symplectic current or through a higher-corner structure in the reduced phase space.

\paragraph{More general D-brane boundary conditions.}
Another natural extension is to analyse more general D-brane boundary conditions. In the present work, the endpoint noncommutativity is governed either by a constant background \(B\)-field or by the effective boundary field strength \(F_{\mu\nu}\). One may generalise this discussion to configurations with different fluxes at the two endpoints, intersecting branes, or spatially varying worldvolume gauge fields. Such systems may lead to inequivalent endpoint symplectic structures and could provide a CPS perspective on more general noncommutative D-brane geometries.

\paragraph{Extension to superstrings.}
It would also be very worthwhile to extend the analysis to superstrings. Since the present paper shows that the tensionless bosonic theory admits a purely boundary supported noncommutative phase space in suitable antisymmetric backgrounds, a supersymmetric generalisation may reveal how fermionic degrees of freedom, supersymmetry, and boundary gauge couplings modify this structure. This could lead to supersymmetric extensions of the endpoint algebra and may clarify whether similar boundary-supported remnants of phase space persist in tensionless superstring theories.

\paragraph{Corner symmetries and endpoint observables.}
The use of the Harlow-Wu framework suggests a broader conceptual direction: one may ask whether the open string endpoints support a nontrivial corner symmetry algebra (may be in the line of \cite{Ciambelli:2021nmv, Ciambelli:2022cfr, Ciambelli:2022vot}). Since the Hamiltonian in the boundary-sensitive CPS formalism receives explicit contributions from corner data, it would be interesting to understand whether the endpoint noncommutativity derived here can be reinterpreted in terms of boundary charges, corner observables, or an emergent endpoint symmetry principle. Such a perspective may provide a deeper geometric understanding of why, in the tensionless regime, the surviving physical phase space is entirely localised at the boundary.


Overall, the CPS approach provides a unified and conceptually transparent description of noncommutativity in open string theory across both the tensile and tensionless regimes. Its central lesson is that the singular tensionless limit does not erase the symplectic structure; rather, it strips it down to its most elemental form, namely a purely boundary-supported noncommutative phase space.

\subsection*{Acknowledgements}
SD is supported by the Shuimu Tsinghua Scholar Program of Tsinghua University and the Beijing Natural Science Foundation of China (Grant No. IS25035). SM thanks the String Theory group of HRI for useful discussions. PKD acknowledges financial support from the DST-INSPIRE fellowship (IF200253), Department of Science and Technology, Govt. of India.

\appendix
\renewcommand{\theequation}{\thesection.\arabic{equation}}

\section{Open string propagator and tensionless limit} \label{sec:open-string_propagator}

In this appendix, we briefly review the worldsheet derivation of the noncommutativity parameter for open strings in a constant $B$-field background \cite{Seiberg:1999vs}, and contrast it with the tensionless limit. 

\subsection*{Open string on the upper half plane}

We consider an open string propagating in the presence of an antisymmetric field $B_{ij}$. The worldsheet is mapped to the upper half plane (UHP), with boundary at $\mathrm{Im}(z)=0$, where
\begin{equation}
z=\tau+i\sigma, \qquad \bar z=\tau-i\sigma.
\end{equation}
The embedding coordinates admit the decomposition
\begin{equation}
x^i(z,\bar z)=X^i(z)+\widetilde X^i(\bar z).
\end{equation}
The presence of the $B$-field modifies the boundary conditions, leading to \cite{Seiberg:1999vs}
\begin{equation}
g_{ij}(\partial-\bar\partial)x^j
+2\pi\alpha' \B_{ij}(\partial+\bar\partial)x^j\Big|_{z=\bar z}=0.
\end{equation}
Here we are using the rescaled Kalb-Ramond field \eqref{eq:rescale_B_tensile}. Restricting to the boundary, this relates the holomorphic and antiholomorphic sectors as
\begin{equation}
(g+2\pi\alpha' \B)_{ij}\,\partial X^j
=(g-2\pi\alpha' \B)_{ij}\,\bar\partial \widetilde X^j.
\end{equation}
Introducing
\begin{equation}
M=(g+2\pi\alpha' \B)^{-1}(g-2\pi\alpha' \B),
\end{equation}
one obtains the reflection relation
\begin{equation}
\widetilde X^i(\bar z)=(M^{-1})^i{}_{\,j}\,X^j(z^*), \qquad z^*=\bar z,
\end{equation}
which allows all correlators on the UHP to be expressed in terms of holomorphic fields.

\subsection*{Worldsheet propagator}

The bulk correlators are given by
\begin{equation}
\langle X^i(z)X^j(w)\rangle=-\alpha' g^{ij}\ln(z-w),
\end{equation}
and similarly for $\widetilde X^i$. Using the reflection relation, the full propagator can be written as a sum over image contributions. A straightforward computation yields
\begin{equation}
\begin{aligned}
\langle x^i(z)x^j(z')\rangle
= -\alpha'\Big[
& g^{ij}\ln(z-z')
+ g^{ik}R^j{}_{\,k}\ln(z-\bar z') \\
& + R^i{}_{\,k}g^{kj}\ln(\bar z-z')
+ R^i{}_{\,k}g^{kl}R^j{}_{\,l}\ln(\bar z-\bar z')
\Big],
\end{aligned}
\end{equation}
where
\begin{equation}
R=(g-2\pi\alpha' \B)^{-1}(g+2\pi\alpha' \B)=M^{-1}.
\end{equation}
It is convenient to separate the symmetric and antisymmetric parts by defining
\begin{equation}
G^{ij}=\tfrac12\big(g^{ik}R^j{}_{\,k}+R^i{}_{\,k}g^{kj}\big), \qquad
\Theta^{ij}=2\pi\alpha'\big(g^{ik}R^j{}_{\,k}-R^i{}_{\,k}g^{kj}\big).
\end{equation}
The propagator then takes the form
\begin{align}
\langle x^i(z)x^j(z')\rangle
&= -\alpha'\Big[ g^{ij}\ln|z-z'| - g^{ij}\ln|z-\bar z'| \nonumber\\
&\quad + G^{ij}\ln|z-\bar z'|^2
+ \tfrac{1}{2\pi\alpha'}\,\Theta^{ij}\ln\!\frac{z-\bar z'}{\bar z - z'} \Big] + D^{ij},
\end{align}
Equivalently, one finds
\begin{equation}
G_{ij}=g_{ij}-(2\pi\alpha')^2(\B g^{-1}\B)_{ij}, \qquad
\Theta^{ij}=2\pi\alpha'\Big(\frac{1}{g+2\pi\alpha' \B}\Big)^{ij}_{A}.
\end{equation}
Here $G_{ij}$ is the effective open-string metric, while $\Theta^{ij}$ encodes the noncommutativity induced by the $B$-field.

\subsection*{Boundary limit and noncommutativity}

Restricting the propagator to the boundary $z=\tau$, $z'=\tau'$, one uses
\begin{equation}
\ln\frac{z-\bar z'}{\bar z-z'}=i\pi\,\epsilon(\tau-\tau'),
\end{equation}
to obtain
\begin{equation}
\langle x^i(\tau)x^j(\tau')\rangle
= -\alpha' G^{ij}\ln(\tau-\tau')^2
+ \frac{i}{2}\Theta^{ij}\epsilon(\tau-\tau').
\end{equation}
The antisymmetric part determines the equal-time commutator
\begin{equation}
[x^i,x^j]=i\,\Theta^{ij},
\end{equation}
which exhibits noncommutativity on the D-brane worldvolume.

\subsection*{Tensionless limit}

For the tensile string, the propagator is logarithmic,
\begin{equation}
\langle X^i(z,\bar z)X^j(w,\bar w)\rangle
= -\frac{\alpha'}{2}g^{ij}\ln|z-w|^2,
\end{equation}
reflecting the underlying two-dimensional conformal symmetry. In the limit $\alpha'\to\infty$, the worldsheet geometry degenerates to a Carrollian structure and the dynamics become ultralocal in the spatial direction.

Choosing the gauge $V^a=(1,0)$, the action reduces to
\begin{equation}
S=\frac{1}{2}\int d\tau d\sigma\, g_{ij}\,\partial_\tau x^i \partial_\tau x^j,
\end{equation}
which contains no spatial derivatives. The corresponding Green’s function satisfies
\begin{equation}
-\partial_\tau^2 \cg_\tau(\tau-\tau')=\delta(\tau-\tau'),
\end{equation}
with solution
\begin{equation}
\cg_\tau(\tau-\tau')=-\frac{1}{2}|\tau-\tau'|.
\end{equation}
The full propagator is therefore
\begin{equation}
\langle x^i(\tau,\sigma)x^j(\tau',\sigma')\rangle
= -\frac{1}{2} g^{ij}|\tau-\tau'|\delta(\sigma-\sigma').
\end{equation}
For periodic $\sigma$, the delta function is replaced by its Fourier representation
\begin{equation}
\delta(\sigma-\sigma')=\frac{1}{2\pi}\sum_{n\in\mathbb{Z}} e^{in(\sigma-\sigma')}.
\end{equation}

Thus, while the tensile theory exhibits a logarithmic propagator governed by conformal symmetry, the tensionless limit leads to a linear, ultralocal propagator characteristic of a Carrollian worldsheet.
\section{Symplectic mechanics}\label{sec:symp_mech}

A symplectic manifold is a pair \( (\Q,\bos{\Omega}) \), where \( \Q \) is a smooth manifold and
\begin{equation}
\bos{\Omega} = \frac{1}{2}\,\Omega_{\mu\nu}(q)\,\hd q^\mu \cw \hd q^\nu
\end{equation}
is a two-form. Here, $\hd$ denotes the exterior derivative defined on the manifold $\Q$, and $\cw$ is the wedge product. The $2$-form $\bos{\Omega}$ satisfies
\begin{equation}
\hd\bos{\Omega} = 0,
\qquad
\det(\Omega_{\mu\nu}) \neq 0.
\end{equation}
The first condition states that \( \bos{\Omega} \) is closed, while the second one states that it is non-degenerate. Non-degeneracy implies that \( \Omega_{\mu\nu} \) is invertible, so there exists a bivector \( \Omega^{\mu\nu} \) such that
\begin{equation}
\Omega^{\mu\alpha}\Omega_{\alpha\nu}=\delta^\mu{}_\nu .
\end{equation}
Since \( \bos{\Omega} \) is closed, it is locally exact. Hence, at least locally, one may write
\begin{equation}
\bos{\Omega} = \hd \bos{\a},
\qquad
\bos{\a} = \a_\mu(q)\,\hd q^\mu ,
\end{equation}
where \( \bos{\a} \) is called the symplectic potential or \emph{tautological $1$-form}\footnote{It is also known as the canonical one-form, the Liouville one-form, or the Poincar\'e one-form depending on the context.} defined on the cotangent bundle $\T^* \Q$. The potential is not unique: it is defined up to the shift
\begin{equation}
\bos{\a} \to \bos{\a} + \hd Z ,
\end{equation}
with \( Z \in C^\infty (\Q) \) . This transformation leaves \( \bos{\Omega}\) unchanged.

\subsection*{Canonical transformations and Hamiltonian vector fields}
Let
\begin{equation}
\hat{\xi}= \hat{\xi}^\mu(q)\,\hdel_\mu
\end{equation}
be a vector field on \( \Q \) with $\{\hdel_\m\}$ being the basis on $\T\Q$. Its action on the symplectic form is determined by Cartan's formula,
\begin{equation}
\mathscr{L}_{\hat{\xi}}\bos{\Omega} = \mathsf{I}_{\hat{\xi}} \hd\bos{\Omega} + \hd(\mathsf{I}_{\hat{\xi}} \bos{\Omega}),
\end{equation}
where $\mathscr{L}_{\hat{\xi}}$ is the Lie derivative defined on $\Q$, and $\mathsf{I}_{\hat{\xi}}$ is the interior derivative (or product) defined as the mapping $\mathsf{I}_{\hat{\xi}} : \Lambda^p \rightarrow \Lambda^{p-1}$ with $\Lambda^p$ being the space of all $p$-forms.
A transformation generated by \( \hat{\xi}\) is canonical if it preserves the symplectic form,
\begin{equation}
\mathscr{L}_{\hat{\xi}} \bos{\Omega} = 0.
\end{equation}
Using \( \hd \bos{\Omega}=0 \), this becomes
\begin{equation}
\hd(\mathsf{I}_{\hat{\xi}} \bos{\Omega})=0.
\end{equation}
Thus, \( \mathsf{I}_{\hat{\xi}}\bos{\Omega} \) is a closed one-form. If the first homology class is trivial, i.e., \( \mathtt{H}^1(\Q)=0 \), every closed one-form is exact, and one may write
\begin{equation}
\mathsf{I}_{\hat{\xi}} \bos{\Omega} = -\hd f
\end{equation}
for some smooth function \( f \) on \( \Q \). The function \( f \) is the Hamiltonian generator of the transformation, and \( \hat{\xi} \) is the corresponding Hamiltonian vector field. In components,
\begin{equation}
\hat{\xi}^\mu = \Omega^{\mu\nu}\,\hdel_\nu f .
\end{equation}
Therefore, the action of the flow generated by \( f \) on another function \( g \) is
\begin{equation}
\hat{\xi}(g)=\Omega^{\mu\nu}\,\hdel_\mu f\,\hdel_\nu g .
\end{equation}
This expression leads directly to the definition of the Poisson bracket. In particular, every smooth function on phase space defines an infinitesimal canonical transformation, and the Hamiltonian function generates time evolution.

\subsection*{Poisson brackets}

For two smooth functions \( f,g \in C^\infty(\Q) \), the Poisson bracket is defined by
\begin{equation}
\{f,g\}=\Omega^{\mu\nu}\,\hdel_\mu f\,\hdel_\nu g .
\end{equation}
Equivalently, it gives the variation of \( g \) under the Hamiltonian flow generated by \( f \). For any observable \( F \),
\begin{equation}
\delta F = \{F,f\}.
\end{equation}
Because \( \Omega^{\mu\nu} \) is antisymmetric, the Poisson bracket is antisymmetric,
\begin{equation}
\{f,g\}=-\{g,f\}.
\end{equation}
Moreover, the closure of \( \bos{\Omega} \) implies the Jacobi identity,
\begin{equation}
\{f,\{g,h\}\}+\{g,\{h,f\}\}+\{h,\{f,g\}\}=0.
\end{equation}
Hence \( C^\infty(\Q) \) forms a Lie algebra under the Poisson bracket. The bracket structure becomes more prominent when we look at the phase space of a theory with canonical coordinates \( (x^i,p_i) \). By using Darboux's theorem, one can always find a local basis for the symplectic form such that it can be expressed as \cite{Nair:2024wyq}
\begin{equation}
\bos{\Omega} = \frac{1}{2}\,\Omega_{\mu\nu}(q)\,\hd q^\mu \cw \hd q^\nu = \hd p_i \cw \hd x^i,
\end{equation}
where $q^\m =(x^i, p_i )$. In these canonical bases, the basic Poisson brackets reduce to
\begin{equation}
\{x^i,x^j\}=0,
\qquad
\{x^i,p_j\}=\delta^i{}_j,
\qquad
\{p_i,p_j\}=0.
\end{equation}
More generally, applying the definition to the coordinates \( q^\mu \) themselves gives
\begin{equation}
\{q^\mu,q^\nu\}=\Omega^{\mu\nu}.
\end{equation}
Thus, the basic Poisson brackets are simply the components of the inverse symplectic form. This provides the direct link between the geometric data \( \Omega_{\mu\nu} \) and the algebraic structure of observables.

\subsection*{The Moyal-Weyl product}

A physically relevant example of noncommutative spaces arises via deformation quantisation of a symplectic manifold, typically applied to the phase space of quantum mechanics or the worldvolume of D-branes \cite{Kontsevich:1997vb, Cornalba:2001sm}. The standard commutative pointwise product of smooth functions $f, g \in C^\infty(\mathbb{R}^{2n})$ is deformed into the Moyal-Weyl star product ($\star$), defined by the pseudo-differential operator:
$$f \star g = f \exp\left( \frac{i}{2}\, \overleftarrow{\pa_\m}\, \Theta^{\m\n} \,\overrightarrow{\pa_\n} \right) g = f \cdot g + \sum_{n=1}^{\infty} \left(\frac{i}{2}\right)^n \frac{1}{n!} \Theta^{\m_1\n_1} \cdots \Theta^{\m_n\n_n} (\pa_{\m_1 \cdots \m_n} f) (\pa_{\n_1 \cdots \n_n} g)$$
where $\Theta^{\mu\nu}$ is a constant real antisymmetric Poisson bivector. This $\star$-product has the following basic properties:

\begin{enumerate}[(i)]
 \item \textit{associativity:} 
 \[f \star (g \star h) = (f \star g) \star h\, ;\]
    
 \item \textit{noncommutativity:} \[f \star g \neq g \star f\,;\]

 \item \textit{cyclicity or trace:} 
 \[\int (f \star g) = \int (g \star f) = \int (f \cdot g)\,.\]
\end{enumerate}
 It is evident that the star product is a deformation of the commutative product with leading order correction proportional to the Poisson bracket
\be
 \{ f, g \}_\Theta = \Theta^{\m\n}\, \pa_\m f \, \pa_\n g \,.
\ee
In particular, the star commutator of the coordinate functions produces the algebraic structure of the spacetime quantisation,
\be
 \left[ x^\m \, \overset{\star}{,} \, x^\n \right] \coloneqq x^\m \star x^\n - x^\n \star x^\m = i \Theta^{\m\n} .
\ee
Kontsevich showed that any Poisson manifold (i.e., a smooth manifold $\Q$ endowed with a Poisson structure $\Theta^{\m\n}$) admits a deformation quantisation \cite{Kontsevich:1997vb}.
\section{Canonical vs covariant phase space} \label{sec:can_vs_cps}

In this appendix, we give a brief overview of the canonical and covariant phase space formalisms. For clarity and brevity, we present the material in a tabular form. Table \ref{tab:phase_space_formal} offers a comparative summary of the canonical phase space and the CPS framework.

\begin{table}[H]
\centering
\renewcommand{\arraystretch}{1.8}
\begin{tabular}{@{} >{\raggedright\arraybackslash}p{3.5cm} >{\raggedright\arraybackslash}p{6cm} >{\raggedright\arraybackslash}p{6cm} @{}}
\toprule
\textbf{Mathematical feature} & \textbf{Canonical phase space} & \textbf{Covariant phase space} \\
\midrule
\textbf{Underlying topology} & Foliation of spacetime $\M \cong \mathbb{R} \times \Sigma$ via the Arnowitt-Deser-Misner (ADM) decomposition. & Manifestly covariant framework on the full spacetime manifold $\mathcal{M}$ without preferential slicing. \\
\textbf{State space geometry} & Space of initial data defined on a spacelike Cauchy hypersurface $\C$. & Space of physical solutions $\til{\P}$ to the Euler-Lagrange equations $\delta S = 0$ over $\M$. \\
\textbf{Fundamental variables} & Fields and conjugate momentum densities $(\phi, \Pi)$, mapping to the cotangent bundle $\T^*\F$. & Field configurations $\phi$ and their linearised variations $\delta\phi$ residing in the tangent space $\T_\phi \til{\P}$. \\
\textbf{Symplectic form} & Non-degenerate symplectic 2-form defined via Darboux coordinates: \(\Omega_{can} = \int_\C \delta \Pi \wedge \delta \phi\) & pre-symplectic 2-form derived from the symplectic current $\omega$: \(\Omega_{cps} = \int_\C \omega(\phi; \d_1\phi, \d_2\phi)\) \\
\textbf{Gauge and constraints} & Managed algebraically via Dirac's formalism; gauge transformations are generated by first-class constraints $\H \approx 0$. & Handled geometrically; gauge transformations correspond to the degenerate directions of $\Omega_{cps}$. The physical phase space is the quotient $\P = \til{\P}/\ker\Omega_{cps}$. \\
\textbf{Conserved charges} & Boundary terms added to the Hamiltonian to ensure differentiability. & Conserved quantities are boundary integrals of the Noether charge $(d-2)$-form $\mathbf{Q}_\xi$. \\
\textbf{Primary advantages} & Suited to well-posed initial value problems (e.g., numerical relativity). & Manifestly preserves full diffeomorphism invariance.  \\
\textbf{Primary drawbacks} & Breaks manifest spacetime covariance, which results in apparent loss of symmetries.  & No canonical structure. \\
\bottomrule
\end{tabular}
\caption{A comparison between the Canonical and Covariant phase space formalisms.}
\label{tab:phase_space_formal}
\end{table}

\section{Alternative derivation with the gauge field}\label{sec:alternative_gauge_derivation}
In the main text, we have seen that including a gauge term for bosonic strings in the presence of a uniform background Kalb-Ramond field naturally fits into the CPSB formalism for computing the symplectic form. Here, we will discuss an alternative method to obtain the same using the Iyer-Wald CPS formalism. We begin with the action \eqref{eq:full-action-bdry}
\begin{equation}
S = -\frac{1}{4\pi\a'} \int_\Sigma d^2\sp \left[ 
\sqrt{-g} g^{ab} \eta_{\mu\nu} \pa_a X^\mu \pa_b X^\nu 
+ \varepsilon^{ab} B_{\m\n} \pa_a X^\mu \pa_b X^\nu \right] 
+ \frac{1}{2\pi\a'} \oint_{\pa\Sigma} d\tau \, A_i(X) \partial_\tau X^i . \nn
\end{equation}
 As mentioned earlier, this action depicts a bosonic string ending on D$p$-brane(s). If both ends of the string are attached to the same brane, one can rewrite the boundary term of the action as \cite{Chu:1998qz, Chu:1999gi},
 \be
 \frac{1}{2\pi\a'} \oint_{\pa\Sigma} d\tau \, A_i(X) \partial_\tau X^i  = - \frac{1}{4\pi\a'} \int_{\Sigma} d^2\sp \, \varepsilon^{ab} F_{\m\n}(X)\, \pa_a X^\m \pa_b X^\n ,
 \ee
where $F_{\m\n}$ is the field strength of the gauge field $A_\m$: $\bos{F}=\ed\bos{A}$. For this special case, the action becomes
\be\label{eq:action_BI}
 S = -\frac{1}{4\pi\a'} \int_\Sigma d^2\sp \left[ 
\sqrt{-g} g^{ab} \eta_{\mu\nu} \pa_a X^\mu \pa_b X^\nu 
+ \varepsilon^{ab} \,\tf_{\m\n} \,\pa_a X^\mu \pa_b X^\nu \right],
\ee
where, $\tf_{\m\n}= B_{\m\n} + F_{\m\n}$ is called the \emph{modified Born-Infeld field strength} tensor.\footnote{Notice that there is a relative sign difference between $\bos{B}$ and $\bos{F}$ here and in \cite{Chu:1998qz,Chu:1999gi}. This difference arises because our string action has an overall minus sign as opposed to the one considered in \cite{Chu:1998qz, Chu:1999gi}.} To obtain the tensionless theory, we first make the replacements:
\begin{align}
-\frac{T}{2}\sqrt{-g} g^{ab} &\longrightarrow V^a V^b, \nn \\
\tf_{\m\n} &\longrightarrow -\frac{\tf_{\m\n}}{\epsilon}, \quad \epsilon \to 0.
\label{eq:gauge_term_replacement}
\end{align}
The vector densities are given by the Lagrange multipliers as defined in \eqref{eq:vector_density_definition}. 
To keep the modified Born-Infeld field strength finite, we introduce a rescaled field strength
\be
 \sf_{\m\n} = \dfrac{1}{2\pi\a'} \tf_{\m\n}.\,
\ee
Replacing this rescaled field and the vector densities in the action \eqref{eq:action_BI}, we obtain the tensionless action 
\begin{equation}
S_{\text{tensionless}} = \int_\Sigma d^2\sp \left( V^a V^b \pa_a X^\m \pa_b X^\n \eta_{\m\n} + \dfrac{1}{2}\veps^{ab} \sf_{\m\n}\, \pa_a X^\m \pa_b X^\n \right) .
\label{eq:tensionless_action_F}
\end{equation}
Now, we can apply the IW method to obtain the symplectic form for the above action. The first term is the tensionless ILST action, which does not contribute to the symplectic form as the symplectic current vanishes identically (see section \ref{sec:cps_tensionless}). The second term can be divided into two parts: one corresponding to the background Kalb-Ramond field, and the other is the field strength of the boundary gauge field. As we have already done the computation for the $\bos{B}$-field (section \ref{sec:cps_tensionless_with_B}), here we will only focus on the field strength in the tensionless limit. The Lagrangian is given by
\be
 \bos{L}_\ff = \dfrac{1}{2} \veps^{ab}\ff_{\m\n} \pa_a X^\m \pa_b X^\n ,
\ee
where $\bos{\ff}$ is the field strength of the boundary gauge field in the tensionless limit, i.e.,  $\bos{\ff} = \ed \bos{\A}$. Variation of the above Lagrangian yields
\begin{align}
   \d  \bos{L}_\ff &= \pa_a \big( \veps^{ab} \ff_{ \m\n} \d X^\m \pa_b X^\n \big) + \pa_a \big( \pa_b \underbrace{(  \A_\n \d X^\n)}_{\mathfrak{c}}\big)  \\
    &= \ed \bost_\ff + \ed (\ed \mathfrak{c}) \,.
\end{align}
The second term vanishes as $\ed^2=0$ by definition, and the symplectic potential is given by
\be
 \theta^a_\ff =  \veps^{ab} \ff_{ \m\n} \, \d X^\m \pa_b X^\n .
\ee
Now, using the definition \eqref{sympcurrent}, we find the symplectic current,
\be
 \omega^a_\ff =   \veps^{ab} \ff_{ \m\n} \, \d X^\m \cw \pa_b \d X^\n  = \pa_b \left( \dfrac{1}{2}\veps^{ab} \ff_{ \m\n} \, \d X^\m \cw \d X^\n \right).
\ee
Combining with the Kalb-Ramond field part, we get the total symplectic current
\be
 \omega^a = \omega^a_B +  \omega^a_\ff = \pa_b \left(  \dfrac{1}{2} \veps^{ab} \sf_{\m\n} \, \d X^\m \cw \d X^\n \right).   
\ee
As done earlier, we can fix $\veps^{\t\sp}=+1$ and write the symplectic form as
\be
 \bos{\Omega}_{\rm{bdry}} =  \dfrac{1}{2} \sf_{\m\n} \, \d X^\m \cw \d X^\n .
\ee
It matches with the symplectic form in section \ref{sec:cpsb_bdry_gauge-field}, where we have used the CPSB method. Thus, when both ends of the string end on the same D$p$-brane, the IW and CPSB methods give us the same symplectic form and hence the same symplectic structure of the boundary phase space.

	

\bibliographystyle{JHEP}
\bibliography{ref}

@article{Isberg:1993av,
    author = "Isberg, J. and Lindstrom, U. and Sundborg, B. and Theodoridis, G.",
    title = "{Classical and quantized tensionless strings}",
    eprint = "hep-th/9307108",
    archivePrefix = "arXiv",
    reportNumber = "USITP-93-12",
    doi = "10.1016/0550-3213(94)90056-6",
    journal = "Nucl. Phys. B",
    volume = "411",
    pages = "122--156",
    year = "1994"
}

@article{Bagchi:2015nca,
    author = "Bagchi, Arjun and Chakrabortty, Shankhadeep and Parekh, Pulastya",
    title = "{Tensionless Strings from Worldsheet Symmetries}",
    eprint = "1507.04361",
    archivePrefix = "arXiv",
    primaryClass = "hep-th",
    reportNumber = "MIT-CTP-4690",
    doi = "10.1007/JHEP01(2016)158",
    journal = "JHEP",
    volume = "01",
    pages = "158",
    year = "2016"
}

@article{Iyer:1994ys,
    author = "Iyer, Vivek and Wald, Robert M.",
    title = "{Some properties of Noether charge and a proposal for dynamical black hole entropy}",
    eprint = "gr-qc/9403028",
    archivePrefix = "arXiv",
    doi = "10.1103/PhysRevD.50.846",
    journal = "Phys. Rev. D",
    volume = "50",
    pages = "846--864",
    year = "1994"
}

@article{Harlow:2019yfa,
    author = "Harlow, Daniel and Wu, Jie-Qiang",
    title = "{Covariant phase space with boundaries}",
    eprint = "1906.08616",
    archivePrefix = "arXiv",
    primaryClass = "hep-th",
    doi = "10.1007/JHEP10(2020)146",
    journal = "JHEP",
    volume = "10",
    pages = "146",
    year = "2020"
}

@article{Hajian:2015xlp,
    author = "Hajian, K. and Sheikh-Jabbari, M. M.",
    title = "{Solution Phase Space and Conserved Charges: A General Formulation for Charges Associated with Exact Symmetries}",
    eprint = "1512.05584",
    archivePrefix = "arXiv",
    primaryClass = "hep-th",
    reportNumber = "IPM-P-2015-073",
    doi = "10.1103/PhysRevD.93.044074",
    journal = "Phys. Rev. D",
    volume = "93",
    number = "4",
    pages = "044074",
    year = "2016"
}

@article{Wald:1993nt,
    author = "Wald, Robert M.",
    title = "{Black hole entropy is the Noether charge}",
    eprint = "gr-qc/9307038",
    archivePrefix = "arXiv",
    reportNumber = "EFI-93-38",
    doi = "10.1103/PhysRevD.48.R3427",
    journal = "Phys. Rev. D",
    volume = "48",
    pages = "R3427--R3431",
    year = "1993"
}

@article{Wald:1999wa,
    author = "Wald, Robert M. and Zoupas, Andreas",
    title = "{A General definition of \"conserved quantities\" in general relativity and other theories of gravity}",
    eprint = "gr-qc/9911095",
    archivePrefix = "arXiv",
    reportNumber = "EFI-99-45",
    doi = "10.1103/PhysRevD.61.084027",
    journal = "Phys. Rev. D",
    volume = "61",
    pages = "084027",
    year = "2000"
}

@article{Crnkovic:1986ex,
    author = "Crnkovic, Cedomir and Witten, Edward",
    title = "{Covariant Description of Canonical Formalism in Geometrical Theories}",
    reportNumber = "Print-86-1309 (PRINCETON)",
    journal = "Print-86-1309 (PRINCETON)",
    month = "9",
    year = "1986"
}

@book{Wald:1984rg,
    author = "Wald, Robert M.",
    title = "{General Relativity}",
    doi = "10.7208/chicago/9780226870373.001.0001",
    publisher = "Chicago Univ. Pr.",
    address = "Chicago, USA",
    year = "1984"
}

@article{Banerjee:2024fbi,
    author = "Banerjee, Aritra and Chatterjee, Ritankar and Pandit, Priyadarshini",
    title = "{Tensionless strings in a Kalb-Ramond background}",
    eprint = "2404.01385",
    archivePrefix = "arXiv",
    primaryClass = "hep-th",
    doi = "10.1007/JHEP06(2024)067",
    journal = "JHEP",
    volume = "06",
    pages = "067",
    year = "2024"
}

@article{Duary:2025hdb,
    author = "Duary, Sarthak and Maji, Sourav",
    title = "{From closed to open strings: the tensionless route in Kalb-Ramond background and noncommutativity}",
    eprint = "2511.20917",
    archivePrefix = "arXiv",
    primaryClass = "hep-th",
    month = "11",
    year = "2025"
}

@article{Bagchi:2026wcu,
    author = "Bagchi, Arjun and Banerjee, Aritra and Chatterjee, Ritankar and Pandit, Priyadarshini",
    title = "{The Tensionless Lives of Null Strings}",
    eprint = "2601.20959",
    archivePrefix = "arXiv",
    primaryClass = "hep-th",
    month = "1",
    year = "2026"
}

@article{Witten:1985cc,
    author = "Witten, Edward",
    title = "{Noncommutative Geometry and String Field Theory}",
    reportNumber = "Print-86-0083 (PRINCETON)",
    doi = "10.1016/0550-3213(86)90155-0",
    journal = "Nucl. Phys. B",
    volume = "268",
    pages = "253--294",
    year = "1986"
}

@article{Seiberg:1999vs,
    author = "Seiberg, Nathan and Witten, Edward",
    title = "{String theory and noncommutative geometry}",
    eprint = "hep-th/9908142",
    archivePrefix = "arXiv",
    reportNumber = "IASSNS-HEP-99-74",
    doi = "10.1088/1126-6708/1999/09/032",
    journal = "JHEP",
    volume = "09",
    pages = "032",
    year = "1999"
}

@article{Sheikh-Jabbari:1997qke,
    author = "Sheikh-Jabbari, M. M.",
    title = "{More on mixed boundary conditions and D-branes bound states}",
    eprint = "hep-th/9712199",
    archivePrefix = "arXiv",
    reportNumber = "IPM-97-260",
    doi = "10.1016/S0370-2693(98)00199-3",
    journal = "Phys. Lett. B",
    volume = "425",
    pages = "48--54",
    year = "1998"
}

@inproceedings{Ardalan:1998ks,
    author = "Ardalan, F. and Arfaei, H. and Sheikh-Jabbari, M. M.",
    title = "{Mixed branes and M(atrix) theory on noncommutative torus}",
    booktitle = "{6th International Symposium on Particles, Strings and Cosmology}",
    eprint = "hep-th/9803067",
    archivePrefix = "arXiv",
    pages = "653--656",
    month = "3",
    year = "1998"
}

@article{Sheikh-Jabbari:1998aur,
    author = "Sheikh-Jabbari, M. M.",
    title = "{SuperYang-Mills theory on noncommutative torus from open strings interactions}",
    eprint = "hep-th/9810179",
    archivePrefix = "arXiv",
    reportNumber = "IPM-P-98-22",
    doi = "10.1016/S0370-2693(99)00122-7",
    journal = "Phys. Lett. B",
    volume = "450",
    pages = "119--125",
    year = "1999"
}

@article{Ardalan:1998ce,
    author = "Ardalan, F. and Arfaei, H. and Sheikh-Jabbari, M. M.",
    title = "{Noncommutative geometry from strings and branes}",
    eprint = "hep-th/9810072",
    archivePrefix = "arXiv",
    reportNumber = "IPM-P-98-19",
    doi = "10.1088/1126-6708/1999/02/016",
    journal = "JHEP",
    volume = "02",
    pages = "016",
    year = "1999"
}

@article{Sheikh-Jabbari:1999krr,
    author = "Sheikh-Jabbari, M. M. and Shirzad, A.",
    title = "{Boundary conditions as Dirac constraints}",
    eprint = "hep-th/9907055",
    archivePrefix = "arXiv",
    reportNumber = "IPM-P-99-037",
    doi = "10.1007/s100520100590",
    journal = "Eur. Phys. J. C",
    volume = "19",
    pages = "383",
    year = "2001"
}

@article{Chu:1999gi,
    author = "Chu, Chong-Sun and Ho, Pei-Ming",
    title = "{Constrained quantization of open string in background B field and noncommutative D-brane}",
    eprint = "hep-th/9906192",
    archivePrefix = "arXiv",
    reportNumber = "NEIP-99-011",
    doi = "10.1016/S0550-3213(99)00685-9",
    journal = "Nucl. Phys. B",
    volume = "568",
    pages = "447--456",
    year = "2000"
}

@article{Chu:1998qz,
    author = "Chu, Chong-Sun and Ho, Pei-Ming",
    title = "{Noncommutative open string and D-brane}",
    eprint = "hep-th/9812219",
    archivePrefix = "arXiv",
    reportNumber = "NEIP-98-022",
    doi = "10.1016/S0550-3213(99)00199-6",
    journal = "Nucl. Phys. B",
    volume = "550",
    pages = "151--168",
    year = "1999"
}

@article{Zuckerman:1986vzu,
    author = "Zuckerman, Gregg J.",
    title = "{ACTION PRINCIPLES AND GLOBAL GEOMETRY}",
    reportNumber = "Print-89-0321 (YALE)",
    journal = "Conf. Proc. C",
    volume = "8607214",
    pages = "259--284",
    year = "1986"
}

@article{Crnkovic:1987tz,
    author = "Crnkovic, Cedomir",
    title = "{Symplectic Geometry of the Covariant Phase Space, Superstrings and Superspace}",
    reportNumber = "PUPT-1059",
    doi = "10.1088/0264-9381/5/12/008",
    journal = "Class. Quant. Grav.",
    volume = "5",
    pages = "1557--1575",
    year = "1988"
}

@article{Lee:1990nz,
    author = "Lee, J. and Wald, Robert M.",
    title = "{Local symmetries and constraints}",
    doi = "10.1063/1.528801",
    journal = "J. Math. Phys.",
    volume = "31",
    pages = "725--743",
    year = "1990"
}

@article{Ashtekar:1990gc,
    author = "Ashtekar, Abhay and Bombelli, Luca and Reula, Oscar",
    title = "{THE COVARIANT PHASE SPACE OF ASYMPTOTICALLY FLAT GRAVITATIONAL FIELDS}",
    reportNumber = "PRINT-90-0318 (SYRACUSE)",
    month = "5",
    year = "1990"
}

@article{Barnich:2007bf,
    author = "Barnich, Glenn and Compere, Geoffrey",
    title = "{Surface charge algebra in gauge theories and thermodynamic integrability}",
    eprint = "0708.2378",
    archivePrefix = "arXiv",
    primaryClass = "gr-qc",
    reportNumber = "ULB-TH-06-30",
    doi = "10.1063/1.2889721",
    journal = "J. Math. Phys.",
    volume = "49",
    pages = "042901",
    year = "2008"
}

@article{Margalef-Bentabol:2022zso,
    author = "Margalef-Bentabol, Juan and Villase{\~n}or, Eduardo J. S.",
    title = "{Proof of the equivalence of the symplectic forms derived from the canonical and the covariant phase space formalisms}",
    eprint = "2204.06383",
    archivePrefix = "arXiv",
    primaryClass = "math-ph",
    doi = "10.1103/PhysRevD.105.L101701",
    journal = "Phys. Rev. D",
    volume = "105",
    number = "10",
    pages = "L101701",
    year = "2022"
}

@article{Margalef-Bentabol:2020teu,
    author = "Margalef-Bentabol, Juan and Villase{\~n}or, Eduardo J. S.",
    title = "{Geometric formulation of the Covariant Phase Space methods with boundaries}",
    eprint = "2008.01842",
    archivePrefix = "arXiv",
    primaryClass = "math-ph",
    doi = "10.1103/PhysRevD.103.025011",
    journal = "Phys. Rev. D",
    volume = "103",
    number = "2",
    pages = "025011",
    year = "2021"
}

@article{Adami:2021nnf,
    author = "Adami, H. and Grumiller, D. and Sheikh-Jabbari, M. M. and Taghiloo, V. and Yavartanoo, H. and Zwikel, C.",
    title = "{Null boundary phase space: slicings, news {\&} memory}",
    eprint = "2110.04218",
    archivePrefix = "arXiv",
    primaryClass = "hep-th",
    doi = "10.1007/JHEP11(2021)155",
    journal = "JHEP",
    volume = "11",
    pages = "155",
    year = "2021"
}

@article{Chandrasekaran:2018aop,
    author = "Chandrasekaran, Venkatesa and Flanagan, {\'E}anna {\'E}. and Prabhu, Kartik",
    title = "{Symmetries and charges of general relativity at null boundaries}",
    eprint = "1807.11499",
    archivePrefix = "arXiv",
    primaryClass = "hep-th",
    doi = "10.1007/JHEP11(2018)125",
    journal = "JHEP",
    volume = "11",
    pages = "125",
    year = "2018",
    note = "[Erratum: JHEP 07, 224 (2023)]"
}

@article{Faulkner:2013ica,
    author = "Faulkner, Thomas and Guica, Monica and Lukyanov, Thomas and Ng, Ling-Yan and Takayanagi, Tadashi",
    title = "{Universal First Law of Entanglement Entropy}",
    eprint = "1312.7856",
    archivePrefix = "arXiv",
    primaryClass = "hep-th",
    doi = "10.1007/JHEP03(2014)051",
    journal = "JHEP",
    volume = "03",
    pages = "051",
    year = "2014"
}

@article{Lashkari:2013koa,
    author = "Lashkari, Nima and McDermott, Michael B. and Van Raamsdonk, Mark",
    title = "{Universal first law of entanglement entropy and circle Virasoro orbit diffs}",
    eprint = "1308.3716",
    archivePrefix = "arXiv",
    primaryClass = "hep-th",
    doi = "10.1007/JHEP04(2014)152",
    journal = "JHEP",
    volume = "04",
    pages = "152",
    year = "2014"
}

@article{Donnelly:2016auv,
    author = "Donnelly, William and Freidel, Laurent",
    title = "{Local subsystems in gauge theory and gravity}",
    eprint = "1601.04744",
    archivePrefix = "arXiv",
    primaryClass = "hep-th",
    doi = "10.1007/JHEP09(2016)102",
    journal = "JHEP",
    volume = "09",
    pages = "102",
    year = "2016"
}

@article{Speranza:2017gxd,
    author = "Speranza, Antony J.",
    title = "{Local emanations of the covariant phase space}",
    eprint = "1706.05061",
    archivePrefix = "arXiv",
    primaryClass = "hep-th",
    doi = "10.1007/JHEP02(2018)021",
    journal = "JHEP",
    volume = "02",
    pages = "021",
    year = "2018"
}

@article{Lashkari:2015hha,
    author = "Lashkari, Nima and Van Raamsdonk, Mark",
    title = "{Canonical Energy is Quantum Fisher Information}",
    eprint = "1508.00897",
    archivePrefix = "arXiv",
    primaryClass = "hep-th",
    doi = "10.1007/JHEP04(2016)153",
    journal = "JHEP",
    volume = "04",
    pages = "153",
    year = "2016"
}

@article{Jafferis:2015del,
    author = "Jafferis, Daniel L. and Lewkowycz, Aitor and Maldacena, Juan and Suh, S. Josephine",
    title = "{Relative entropy equals bulk relative entropy}",
    eprint = "1512.06431",
    archivePrefix = "arXiv",
    primaryClass = "hep-th",
    doi = "10.1007/JHEP06(2016)004",
    journal = "JHEP",
    volume = "06",
    pages = "004",
    year = "2016"
}

@inproceedings{Anderson1992IntroductionTT,
  title={Introduction to the Variational Bicomplex},
  author={Ian Anderson},
  year={1992},
  url={https://api.semanticscholar.org/CorpusID:55453884}
}

@article{Kontsevich:1997vb,
    author = "Kontsevich, Maxim",
    title = "{Deformation quantization of Poisson manifolds. 1.}",
    eprint = "q-alg/9709040",
    archivePrefix = "arXiv",
    doi = "10.1023/B:MATH.0000027508.00421.bf",
    journal = "Lett. Math. Phys.",
    volume = "66",
    pages = "157--216",
    year = "2003"
}

@article{Cornalba:2001sm,
    author = "Cornalba, Lorenzo and Schiappa, Ricardo",
    title = "{Nonassociative star product deformations for D-brane world volumes in curved backgrounds}",
    eprint = "hep-th/0101219",
    archivePrefix = "arXiv",
    reportNumber = "LPTENS-01-04, HUTP-01-A002",
    doi = "10.1007/s002201000569",
    journal = "Commun. Math. Phys.",
    volume = "225",
    pages = "33--66",
    year = "2002"
}

@article{Hollands:2012sf,
    author = "Hollands, Stefan and Wald, Robert M.",
    title = "{Stability of Black Holes and Black Branes}",
    eprint = "1201.0463",
    archivePrefix = "arXiv",
    primaryClass = "gr-qc",
    doi = "10.1007/s00220-012-1638-1",
    journal = "Commun. Math. Phys.",
    volume = "321",
    pages = "629--680",
    year = "2013"
}

@article{Hollands:2024vbe,
    author = "Hollands, Stefan and Wald, Robert M. and Zhang, Victor G.",
    title = "{Entropy of dynamical black holes}",
    eprint = "2402.00818",
    archivePrefix = "arXiv",
    primaryClass = "hep-th",
    doi = "10.1103/PhysRevD.110.024070",
    journal = "Phys. Rev. D",
    volume = "110",
    number = "2",
    pages = "024070",
    year = "2024"
}

@article{Wall:2015raa,
    author = "Wall, Aron C.",
    title = "{A Second Law for Higher Curvature Gravity}",
    eprint = "1504.08040",
    archivePrefix = "arXiv",
    primaryClass = "gr-qc",
    doi = "10.1142/S0218271815440149",
    journal = "Int. J. Mod. Phys. D",
    volume = "24",
    number = "12",
    pages = "1544014",
    year = "2015"
}

@article{Faulkner:2017tkh,
    author = "Faulkner, Thomas and Haehl, Felix M. and Hijano, Eliot and Parrikar, Onkar and Rabideau, Charles and Van Raamsdonk, Mark",
    title = "{Nonlinear Gravity from Entanglement in Conformal Field Theories}",
    eprint = "1705.03026",
    archivePrefix = "arXiv",
    primaryClass = "hep-th",
    doi = "10.1007/JHEP08(2017)057",
    journal = "JHEP",
    volume = "08",
    pages = "057",
    year = "2017"
}

@article{Haehl:2017sot,
    author = "Haehl, Felix M. and Hijano, Eliot and Parrikar, Onkar and Rabideau, Charles",
    title = "{Higher Curvature Gravity from Entanglement in Conformal Field Theories}",
    eprint = "1712.06620",
    archivePrefix = "arXiv",
    primaryClass = "hep-th",
    doi = "10.1103/PhysRevLett.120.201602",
    journal = "Phys. Rev. Lett.",
    volume = "120",
    number = "20",
    pages = "201602",
    year = "2018"
}

@article{Das:2025fav,
    author = "Das, Pratik K. and Mahato, Manavendra",
    title = "{Bit threads: from entanglement to geometric entropies}",
    eprint = "2508.18941",
    archivePrefix = "arXiv",
    primaryClass = "hep-th",
    doi = "10.1007/JHEP01(2026)049",
    journal = "JHEP",
    volume = "01",
    pages = "049",
    year = "2026"
}

@article{Agon:2020mvu,
    author = "Ag{\'o}n, Cesar A. and C{\'a}ceres, Elena and Pedraza, Juan F.",
    title = "{Bit threads, Einstein{\textquoteright}s equations and bulk locality}",
    eprint = "2007.07907",
    archivePrefix = "arXiv",
    primaryClass = "hep-th",
    reportNumber = "YITP-20-19, UTTG-04-2020, BRX-TH-6666",
    doi = "10.1007/JHEP01(2021)193",
    journal = "JHEP",
    volume = "01",
    pages = "193",
    year = "2021"
}

@article{Parvizi:2025wsg,
    author = "Parvizi, A. and Sheikh-Jabbari, Mohammad M. and Taghiloo, Vahid",
    title = "{Freelance holography, part II: Moving boundary in gauge/gravity correspondence}",
    eprint = "2503.09372",
    archivePrefix = "arXiv",
    primaryClass = "hep-th",
    doi = "10.21468/SciPostPhysCore.8.4.075",
    journal = "SciPost Phys. Core",
    volume = "8",
    pages = "075",
    year = "2025"
}

@article{Parvizi:2025shq,
    author = "Parvizi, A. and Sheikh-Jabbari, Mohammad M. and Taghiloo, Vahid",
    title = "{Freelance holography, part I: Setting boundary conditions free in gauge/gravity correspondence}",
    eprint = "2503.09371",
    archivePrefix = "arXiv",
    primaryClass = "hep-th",
    doi = "10.21468/SciPostPhys.19.2.043",
    journal = "SciPost Phys.",
    volume = "19",
    number = "2",
    pages = "043",
    year = "2025"
}

@article{Visser:2024pwz,
    author = "Visser, Manus R. and Yan, Zihan",
    title = "{Properties of dynamical black hole entropy}",
    eprint = "2403.07140",
    archivePrefix = "arXiv",
    primaryClass = "hep-th",
    doi = "10.1007/JHEP10(2024)029",
    journal = "JHEP",
    volume = "10",
    pages = "029",
    year = "2024"
}

@article{Chandrasekaran:2021vyu,
    author = "Chandrasekaran, Venkatesa and Flanagan, Eanna E. and Shehzad, Ibrahim and Speranza, Antony J.",
    title = "{A general framework for gravitational charges and holographic renormalization}",
    eprint = "2111.11974",
    archivePrefix = "arXiv",
    primaryClass = "gr-qc",
    doi = "10.1142/S0217751X22501056",
    journal = "Int. J. Mod. Phys. A",
    volume = "37",
    number = "17",
    pages = "2250105",
    year = "2022"
}

@article{Ciambelli:2021nmv,
    author = "Ciambelli, Luca and Leigh, Robert G. and Pai, Pin-Chun",
    title = "{Embeddings and Integrable Charges for Extended Corner Symmetry}",
    eprint = "2111.13181",
    archivePrefix = "arXiv",
    primaryClass = "hep-th",
    doi = "10.1103/PhysRevLett.128.171302",
    journal = "Phys. Rev. Lett.",
    volume = "128",
    year = "2022"
}

@article{Ciambelli:2022cfr,
    author = "Ciambelli, Luca and Leigh, Robert G.",
    title = "{Universal corner symmetry and the orbit method for gravity}",
    eprint = "2207.06441",
    archivePrefix = "arXiv",
    primaryClass = "hep-th",
    doi = "10.1016/j.nuclphysb.2022.116053",
    journal = "Nucl. Phys. B",
    volume = "986",
    pages = "116053",
    year = "2023"
}

@article{Ciambelli:2022vot,
    author = "Ciambelli, Luca",
    title = "{From Asymptotic Symmetries to the Corner Proposal}",
    eprint = "2212.13644",
    archivePrefix = "arXiv",
    primaryClass = "hep-th",
    doi = "10.22323/1.435.0002",
    journal = "PoS",
    volume = "Modave2022",
    pages = "002",
    year = "2023"
}

@article{Cho:2023khj,
    author = "Cho, Minjae and Mazel, Ben and Yin, Xi",
    title = "{Rolling tachyon and the phase space of open string field theory}",
    eprint = "2310.17895",
    archivePrefix = "arXiv",
    primaryClass = "hep-th",
    doi = "10.1007/JHEP04(2025)129",
    journal = "JHEP",
    volume = "04",
    pages = "129",
    year = "2025"
}

@article{Bernardes:2025uzg,
    author = "Bernardes, Vin{\'\i}cius and Erler, Theodore and F{\i}rat, Atakan Hilmi",
    title = "{Covariant phase space and L$_{\infty}$ algebras}",
    eprint = "2506.20706",
    archivePrefix = "arXiv",
    primaryClass = "hep-th",
    reportNumber = "MIT-CTP-5880",
    doi = "10.1007/JHEP09(2025)057",
    journal = "JHEP",
    volume = "09",
    pages = "057",
    year = "2025"
}

@article{Bernardes:2025zzu,
    author = "Bernardes, Vin{\'\i}cius and Erler, Theodore and F{\i}rat, Atakan Hilmi",
    title = "{Symplectic structure in open string field theory. Part I. Rolling tachyons}",
    eprint = "2511.03777",
    archivePrefix = "arXiv",
    primaryClass = "hep-th",
    doi = "10.1007/JHEP02(2026)063",
    journal = "JHEP",
    volume = "02",
    pages = "063",
    year = "2026"
}

@article{Bernardes:2025zkj,
    author = "Bernardes, Vin{\'\i}cius and Erler, Theodore and F{\i}rat, Atakan Hilmi",
    title = "{Symplectic structure in open string field theory. Part II. Sliding lump}",
    eprint = "2511.15781",
    archivePrefix = "arXiv",
    primaryClass = "hep-th",
    doi = "10.1007/JHEP02(2026)064",
    journal = "JHEP",
    volume = "02",
    pages = "064",
    year = "2026"
}

@article{Bernardes:2026egp,
    author = "Bernardes, Vin{\'\i}cius and Erler, Theodore and F{\i}rat, Atakan Hilmi",
    title = "{Symplectic structure in open string field theory III: Electric field}",
    eprint = "2604.01273",
    archivePrefix = "arXiv",
    primaryClass = "hep-th",
    month = "4",
    year = "2026"
}

@article{Barnich:2001jy,
    author = "Barnich, Glenn and Brandt, Friedemann",
    title = "{Covariant theory of asymptotic symmetries, conservation laws and central charges}",
    eprint = "hep-th/0111246",
    archivePrefix = "arXiv",
    reportNumber = "ULB-TH-01-19, MPI-MIS-94-2001",
    doi = "10.1016/S0550-3213(02)00251-1",
    journal = "Nucl. Phys. B",
    volume = "633",
    pages = "3--82",
    year = "2002"
}

@article{Speziale:2025lkm,
    author = "Speziale, Simone",
    title = "{GGI lectures on boundary and asymptotic symmetries}",
    eprint = "2512.16810",
    archivePrefix = "arXiv",
    primaryClass = "hep-th",
    month = "12",
    year = "2025"
}

@book{Green:1987sp,
    author = "Green, Michael B. and Schwarz, J. H. and Witten, Edward",
    title = "{SUPERSTRING THEORY. VOL. 1: INTRODUCTION}",
    isbn = "978-0-521-35752-4",
    series = "Cambridge Monographs on Mathematical Physics",
    month = "7",
    year = "1988"
}

@book{Nair:2024wyq,
    author = "Nair, V. Parameswaran",
    title = "{Geometric Quantization and Applications to Fields and Fluids}",
    doi = "10.1007/978-3-031-65801-3",
    isbn = "978-3-031-65801-3, 978-3-031-65800-6",
    publisher = "Springer Nature",
    year = "2024"
}

@article{Chandrasekaran:2026pnc,
    author = "Chandrasekaran, Venkatesa and Flanagan, {\'E}anna {\'E}.",
    title = "{Subregion algebras in classical and quantum gravity}",
    eprint = "2601.07915",
    archivePrefix = "arXiv",
    primaryClass = "hep-th",
    month = "1",
    year = "2026"
}

@article{Ali:2026kmk,
    author = "Ali, Mohd and Stettinger, Georg",
    title = "{The BEF Symplectic Form: A Lagrangian Perspective}",
    eprint = "2604.07334",
    archivePrefix = "arXiv",
    primaryClass = "hep-th",
    month = "4",
    year = "2026"
}

@article{Blumenhagen:2014sba,
    author = "Blumenhagen, Ralph",
    editor = {Anagnostopoulos, Konstantinos and Lechtenfeld, Olaf and L{\"u}st, Dieter and Kawai, Hikaru and Nishimura, Jun and Steinacker, Harold and Szabo, Richard and Watamura, Satoshi and Zoupanos, George},
    title = "{A Course on Noncommutative Geometry in String Theory}",
    eprint = "1403.4805",
    archivePrefix = "arXiv",
    primaryClass = "hep-th",
    reportNumber = "MPP-2014-54",
    doi = "10.1002/prop.201400014",
    journal = "Fortsch. Phys.",
    volume = "62",
    pages = "709--726",
    year = "2014"
}

@article{Schomerus:1999ug,
    author = "Schomerus, Volker",
    title = "{D-branes and deformation quantization}",
    eprint = "hep-th/9903205",
    archivePrefix = "arXiv",
    reportNumber = "DESY-99-039, MLI-13-1998-99",
    doi = "10.1088/1126-6708/1999/06/030",
    journal = "JHEP",
    volume = "06",
    pages = "030",
    year = "1999"
}

@article{Herbst:2003we,
    author = "Herbst, Manfred and Kling, Alexander and Kreuzer, Maximilian",
    title = "{Cyclicity of nonassociative products on D-branes}",
    eprint = "hep-th/0312043",
    archivePrefix = "arXiv",
    reportNumber = "TUW-03-33, ITP-UH-32-03, CERN-TH-2003-280",
    doi = "10.1088/1126-6708/2004/03/003",
    journal = "JHEP",
    volume = "03",
    pages = "003",
    year = "2004"
}

@article{Herbst:2001ai,
    author = "Herbst, Manfred and Kling, Alexander and Kreuzer, Maximilian",
    title = "{Star products from open strings in curved backgrounds}",
    eprint = "hep-th/0106159",
    archivePrefix = "arXiv",
    reportNumber = "TUW-01-18",
    doi = "10.1088/1126-6708/2001/09/014",
    journal = "JHEP",
    volume = "09",
    pages = "014",
    year = "2001"
}

@article{Connes:1997cr,
    author = "Connes, Alain and Douglas, Michael R. and Schwarz, Albert S.",
    title = "{Noncommutative geometry and matrix theory: Compactification on tori}",
    eprint = "hep-th/9711162",
    archivePrefix = "arXiv",
    reportNumber = "RU-97-94",
    doi = "10.1088/1126-6708/1998/02/003",
    journal = "JHEP",
    volume = "02",
    pages = "003",
    year = "1998"
}

@article{Douglas:1997fm,
    author = "Douglas, Michael R. and Hull, Christopher M.",
    title = "{D-branes and the noncommutative torus}",
    eprint = "hep-th/9711165",
    archivePrefix = "arXiv",
    reportNumber = "IHES-P-97-83, RU-97-95, QMW-PH-97-34, LPTENS-97-56",
    doi = "10.1088/1126-6708/1998/02/008",
    journal = "JHEP",
    volume = "02",
    pages = "008",
    year = "1998"
}

@article{Schwarz:1998qj,
    author = "Schwarz, Albert S.",
    title = "{Morita equivalence and duality}",
    eprint = "hep-th/9805034",
    archivePrefix = "arXiv",
    reportNumber = "UCD-9805",
    doi = "10.1016/S0550-3213(98)00550-1",
    journal = "Nucl. Phys. B",
    volume = "534",
    pages = "720--738",
    year = "1998"
}

@article{Rieffel:1998vs,
    author = "Rieffel, Marc A. and Schwarz, Albert S.",
    title = "{Morita equivalence of multidimensional noncommutative tori}",
    eprint = "math/9803057",
    archivePrefix = "arXiv",
    doi = "10.1142/S0129167X99000100",
    journal = "Int. J. Math.",
    volume = "10",
    pages = "289--299",
    year = "1999"
}

@article{Astashkevich:1998uc,
    author = "Astashkevich, Alexander and Nekrasov, Nikita and Schwarz, Albert S.",
    title = "{On noncommutative Nahm transform}",
    eprint = "hep-th/9810147",
    archivePrefix = "arXiv",
    reportNumber = "HUTP--98-A054, ITEP-TH-62-98",
    doi = "10.1007/s002200050807",
    journal = "Commun. Math. Phys.",
    volume = "211",
    pages = "167--182",
    year = "2000"
}

@inproceedings{Morariu:1998qm,
    author = "Morariu, Bogdan and Zumino, Bruno",
    title = "{SuperYang-Mills on the noncommutative torus}",
    booktitle = "{Richard Arnowitt Fest: A Symposium on Supersymmetry and Gravitation}",
    eprint = "hep-th/9807198",
    archivePrefix = "arXiv",
    reportNumber = "UCB-PTH-98-38, LBNL-42104, LBL-42104",
    pages = "53--69",
    month = "7",
    year = "1998"
}

@article{Brace:1998ai,
    author = "Brace, Daniel and Morariu, Bogdan",
    title = "{A Note on the BPS spectrum of the matrix model}",
    eprint = "hep-th/9810185",
    archivePrefix = "arXiv",
    reportNumber = "UCB-PTH-98-51, LBNL-42423, LBL-42423",
    doi = "10.1088/1126-6708/1999/02/004",
    journal = "JHEP",
    volume = "02",
    pages = "004",
    year = "1999"
}

@article{Brace:1998ku,
    author = "Brace, Daniel and Morariu, Bogdan and Zumino, Bruno",
    title = "{Dualities of the matrix model from T duality of the Type II string}",
    eprint = "hep-th/9810099",
    archivePrefix = "arXiv",
    reportNumber = "UCB-PTH-98-48, LBL-42383, LBNL-42383",
    doi = "10.1016/S0550-3213(99)00009-7",
    journal = "Nucl. Phys. B",
    volume = "545",
    pages = "192--216",
    year = "1999"
}

@article{Brace:1998xz,
    author = "Brace, Daniel and Morariu, Bogdan and Zumino, Bruno",
    title = "{T duality and Ramond-Ramond backgrounds in the matrix model}",
    eprint = "hep-th/9811213",
    archivePrefix = "arXiv",
    reportNumber = "UCB-PTH-98-56, LBNL-42548, LBL-42548",
    doi = "10.1016/S0550-3213(99)00168-6",
    journal = "Nucl. Phys. B",
    volume = "549",
    pages = "181--193",
    year = "1999"
}

@article{Hofman:1998iy,
    author = "Hofman, Christiaan and Verlinde, Erik P.",
    title = "{U duality of Born-Infeld on the noncommutative two torus}",
    eprint = "hep-th/9810116",
    archivePrefix = "arXiv",
    reportNumber = "SPIN-1998-02",
    doi = "10.1088/1126-6708/1998/12/010",
    journal = "JHEP",
    volume = "12",
    pages = "010",
    year = "1998"
}

@article{Nekrasov:1998ss,
    author = "Nekrasov, Nikita and Schwarz, Albert S.",
    title = "{Instantons on noncommutative R**4 and (2,0) superconformal six-dimensional theory}",
    eprint = "hep-th/9802068",
    archivePrefix = "arXiv",
    reportNumber = "ITEP-TH-9-98, HUTP-98-A004, DAVIS-02-98",
    doi = "10.1007/s002200050490",
    journal = "Commun. Math. Phys.",
    volume = "198",
    pages = "689--703",
    year = "1998"
}

@article{Berkooz:1998st,
    author = "Berkooz, Micha",
    title = "{Nonlocal field theories and the noncommutative torus}",
    eprint = "hep-th/9802069",
    archivePrefix = "arXiv",
    reportNumber = "IASSNS-HEP-98-12, NSF-ITP-98-018",
    doi = "10.1016/S0370-2693(98)00410-9",
    journal = "Phys. Lett. B",
    volume = "430",
    pages = "237--241",
    year = "1998"
}

@article{Reyes:2004zz,
    author = "Reyes, E. G.",
    title = "{On Covariant Phase Space and the Variational Bicomplex}",
    doi = "10.1023/B:IJTP.0000048614.90426.2f",
    journal = "Int. J. Theor. Phys.",
    volume = "43",
    pages = "1267--1286",
    year = "2004"
}

\end{document}